\documentclass{emulateapj}
 \RequirePackage{lineno}
 \usepackage{amsmath}
 \usepackage{amssymb}
 \usepackage{amsthm}
 \usepackage{natbib, url,bm}
 \usepackage{array}
 \usepackage{float}
 \usepackage{graphicx}
 \usepackage{subfigure}
 \usepackage{color}

\begin{document}

\title{Can Winds Driven by Active Galactic Nuclei Account for \\
the Extragalactic Gamma-ray and Neutrino Backgrounds?}

\author{Ruo-Yu Liu\altaffilmark{1,7}, Kohta Murase\altaffilmark{2,3}, Susumu Inoue\altaffilmark{4}, Chong Ge\altaffilmark{5}, Xiang-Yu Wang\altaffilmark{6}}
\altaffiltext{1}{Max-Planck-Institut f\"ur Kernphysik, D-69117
Heidelberg, Germany} 
\altaffiltext{2}{Department of Physics; Department of Astronomy \& Astrophysics; Center for Particle and Gravitational Astrophysics, The Pennsylvania State University, University Park, Pennsylvania 16802, USA}
\altaffiltext{3}{Yukawa Institute for Theoretical Physics, Kyoto, Kyoto, 606-8502, Japan}
\altaffiltext{4}{Astrophysical Big Bang Laboratory, RIKEN, 2-1 Hirosawa, Wako 351-0198 Japan}
\altaffiltext{5}{Department of Physics, University of Alabama in Huntsville, Huntsville, AL 35899, USA}
\altaffiltext{6}{School of Astronomy and Space Science, Nanjing University, Nanjing 210093, China} 
\altaffiltext{7}{Deutsches Elektronen-Synchrotron (DESY), Platanenallee 6, D-15738 Zeuthen, Germany; ruoyu.liu@desy.de}

\begin{abstract}
Various observations are revealing the widespread occurrence of fast and powerful winds in active galactic nuclei (AGNs) that are distinct from relativistic jets, likely launched from accretion disks and interacting strongly with the gas of their host galaxies. During the interaction, strong shocks are expected to form that can accelerate nonthermal particles to high energies. Such winds have been suggested to be responsible for a large fraction of the observed extragalactic gamma-ray background (EGB) and the diffuse neutrino background, via the decay of neutral and charged pions generated in inelastic $pp$ collisions between protons accelerated by the forward shock and the ambient gas. However, previous studies did not properly account for processes such as adiabatic losses that may reduce the gamma-ray and neutrino fluxes significantly. We evaluate the production of gamma rays and neutrinos by AGN-driven winds in detail by modeling their hydrodynamic and thermal evolution, including the effects of their two-temperature structure. We find that they can only account for less than $\sim 30$\% of the EGB flux, as otherwise the model would violate the independent upper limit derived from the diffuse isotropic gamma-ray background. If the neutrino spectral index is steep with $\Gamma\gtrsim 2.2$, a severe tension with the isotropic gamma-ray background would arise as long as the winds contribute more than $20$\% of the IceCube neutrino flux in the $10-100$\,TeV range. At energies $\gtrsim100$~TeV, we find that the IceCube neutrino flux may still be accountable by AGN-driven winds if the spectral index is as small as $\Gamma\sim2.0-2.1$.
\end{abstract}

\section{Introduction}
Powerful, highly collimated jets of predominantly nonthermal plasma with ultrarelativistic outflow velocities are seen to be produced in less than 10\% of all active galactic nuclei (AGNs; e.g. \citealt{Peterson97}). On the other hand, there is widespread evidence that AGNs can more commonly eject moderately collimated winds of thermal plasma, with outflow velocities from a few thousand kilometers per second up to mildly relativistic values of $\sim 0.3c$ (where $c$ is the speed of light), primarily observed as blue-shifted absorption features due to ionized metals at UV and X-ray energies in Seyfert galaxies and quasars \citep[for reviews, see, e.g.][]{Crenshaw03, Veilleux05, Fabian12, King15, Tombesi16}. The winds are inferred to be generated on subparsec scales, and their estimated kinetic power can reach a fair fraction of the AGN bolometric luminosity. Recent observations of fast and massive outflows of atomic and molecular material in AGNs are likely evidence that such winds propagate to kiloparsec scales and interact strongly with the gas in their host galaxies \citep[e.g.][]{Cicone14, Tombesi15}.

Such AGN-driven winds may be launched from accretion disks by a variety of mechanisms involving thermal, radiative and magnetic processes \citep{Crenshaw03, Ohsuga14}. They may be the primary agents by which supermassive black holes (SMBHs) provide mechanical and/or thermal feedback onto their host galaxies, potentially leading to the observed black hole scaling relations and/or the quenching of star formation in massive galaxies \citep[for reviews, see e.g.][]{Fabian12, Kormendy13, Heckman14, King15}. Some recent theoretical studies have addressed the interaction of AGN-driven winds with the ambient gas in the host galaxy and/or halo, with particular attention to the physics of the resulting shocks and its observational consequences \citep{JiangYF10, FG12, Nims15, WL15}.

The extragalactic gamma-ray background (EGB) has been measured in the GeV--TeV range by the Fermi Large Area Telescope (Fermi-LAT, \citealt{Fermi15}), and the diffuse neutrino intensity at $\gtrsim 10\,$TeV energies has been observed by the IceCube  Neutrino Observatory \citep[e.g.,][]{IC13_sci,IC13_prl,IC15,Halzen17}. 
Astrophysical gamma rays and neutrinos can be produced via the decay of neutral ($\pi^0\to \gamma\gamma$) and charged pions ($\pi^\pm \to e^\pm\nu_\mu\nu_e$), and one of the main meson production mechanisms is the hadronuclear reaction (or inelastic $pp$ collision) between high-energy cosmic-rays and cold nucleons in the ambient medium. It has been suggested that when the AGN-driven wind expands into the ISM of the host galaxy, cosmic rays with nuclear charge number $Z$ are accelerated up to $\sim Z \times (10^{16} -10^{17})$ eV by the forward shock \citep{Tamborra14,WL16a, WL16b, Lamastra17}. \cite{Tamborra14} \citep[see also][]{Murase14} pointed out that the contributions of starburst galaxies coexisting with AGNs are necessary for star-forming galaxies to significantly contribute to the diffuse neutrino and gamma-ray backgrounds, and suggested the possibility of AGN-driven winds as one of the cosmic-ray accelerators. However, realistically, the theoretical gamma-ray and neutrino fluxes highly depend on the model parameters, such as the shock velocity evolution and the density of the ambient medium which determines the interaction efficiency, as studied in \cite{WL16a, WL16b, Lamastra17}. { Actually, as we will show in this work, the total diffuse neutrino background and EGB can not be simultaneously explained by this model, once considering the constraint from the so-called isotropic gamma-ray background (IGRB), which is obtained by subtracting the emission of resolved extragalactic point sources from the EGB \citep{Fermi15}.}

In this work, we evaluate the gamma-ray and neutrino emission from AGN-driven winds in more detail compared to previous studies. We take into account several effects that had not been properly accounted for, such as the two-temperature structure of the wind, and the adiabatic cooling of accelerated protons. The resulting diffuse gamma-ray and neutrino fluxes are reduced, by which we can avoid the problem of overshooting the IGRB. The paper is structured as follows: the dynamical evolution of the wind is studied in Section 2; gamma-ray and neutrino production by an individual source is calculated in Section 3; we obtain the diffuse gamma-ray and neutrino flux from the sources throughout the universe and compare with the results in previous literatures in Section 4; in Section 5 and Section 6, we discuss various implications of our results and the summary is given in Section 7.

\section{Dynamics of AGN-driven Winds}
\begin{figure}[htbp]
\centering
\includegraphics[width=0.9\columnwidth]{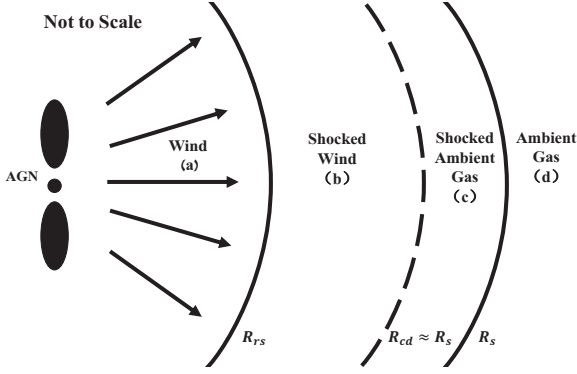}
\caption{Schematic diagram for the structure of the AGN wind-shock system. Spherical symmetry is assumed for the system. See text for detailed descriptions.}\label{fig:sd}
\end{figure}

Following \citet[][hereafter, WLI and WLII]{WL16a, WL16b} and \citet{Lamastra17}, we adopt the 1D model and assume the spherical symmetry for the wind and the ambient gas. The physical picture is similar to that of the stellar-wind bubble \citep{Castor75} but in different scales. 
Let us denote the radius of the forward shock that expands into the ambient medium by $R_{\rm s}$, and the radius of the reverse shock which decelerates the wind by $R_{\rm rs}$. Together with a contact discontinuity at radius $R_{\rm cd}$ which separates the two shocks, this dynamical system are divided into four distinct zones. Outward, they are: (a) the cold fast AGN-wind moving with the bulk velocity $v_{\rm w}$; (b) the hot shocked winds; (c) the shocked ambient gases and (d) the ambient gas which are assumed to consist of pure hydrogen atoms for simplicity. { A schematic diagram which illustrates the outflow structure is shown in Fig.~\ref{fig:sd}.} 
Following the treatment in the previous literatures \citep{Weaver77, FG12, WL15}, we consider the so-called thin shell approximation for region~c which assumes negligible thickness of the shocked ambient gases (i.e., $R_{cd}\simeq R_s$) and all the shocked gases move with the same velocity $v_{\rm s}$\footnote{The forward shock speed should be about 4/3 times the downstream speed when the Mach number is large. But they are essentially the same under the thin-shell approximation.}. In region~b or the region of shocked AGN wind, we consider a steady flow of a homogeneous density $n_{\rm sw}$ and temperature $T_{\rm sw}$ which results in a homogeneous thermal pressure $P_{\rm sw}$ in the region at any given time. The condition of mass conservation then gives a constant value of $R^2v_{\rm sw}$ from $R_{\rm rs}$ to $R_{s}$ where $R$ is the distance to the AGN at the galactic center and $v_{\rm sw}$ is the velocity of the shocked wind. At $R=R_{s}$, the shocked wind should move as the same velocity of the shocked gas, so we have the boundary condition, $v_{\rm sw}(R_{\rm s})=v_{\rm s}$. 
Let's further denote the velocity of the shocked wind just behind the reverse shock by $v_{\rm sw}(R_{\rm rs})=v_{\rm sw}'$, and then we have $v_{\rm sw}'=(R_s/R_{\rm rs})^2v_s$. We note that the velocity of the shocked wind just behind the reverse shock is not equal to that of the reverse shock $v_{\rm rs}$. But we can find the relation between the them by the Rankine-Hugoniot jump relation, i.e.,
\begin{equation}\label{eq:rh_v}
v_w-v_{\rm rs}=4(v_{\rm sw}'-v_{\rm rs}),
\end{equation}
Besides, this condition gives the proton and electron temperatures in shocked wind immediately behind the shock by 
\begin{eqnarray}
T_{p,\rm sw}=\frac{3}{16}\frac{m_p}{k}(v_{\rm w}-v_{\rm rs})^2,\label{eq:rh_tem_p}\\
T_{e,\rm sw}=\frac{3}{16}\frac{m_e}{k}(v_{\rm w}-v_{\rm rs})^2. \label{eq:rh_tem_e}
\end{eqnarray}
{where $m_p$ and $m_e$ are the mass of a proton and an electron, respectively}. We consider the minimal electron heating case and protons receive the majority of the shock heat \citep{FG12}, and the thermal pressure of the shocked wind can then be found by
\begin{eqnarray}\label{eq:rh_P}
\begin{split}
P_{p,\rm sw}&=n_{\rm sw}kT_{\rm sw}=\frac{3}{16}n_{\rm sw}m_p(v_{\rm w}-v_{\rm rs})^2\\
&=\frac{1}{3}n_{\rm sw}m_p(v_{\rm w}-v_{\rm sw}')^2,
\end{split}\\
\begin{split}
P_{e,\rm sw}&=n_{\rm sw}kT_{\rm sw}=\frac{3}{16}n_{\rm sw}m_e(v_{\rm w}-v_{\rm rs})^2\\
&=\frac{1}{3}n_{\rm sw}m_e(v_{\rm w}-v_{\rm sw}')^2,
\end{split}
\end{eqnarray} 
and the total thermal pressure is $P_{\rm sw}=P_{p,\rm sw}+P_{e,\rm sw}$. In the above expressions, $n_{\rm sw}=4n_{\rm w}=\frac{\dot{M}_{\rm w}}{\pi R_{\rm rs}^2m_pv_{\rm w}}$ is the density of both protons and electrons in the shocked wind, where $n_{\rm w}$ is the density of the unshocked wind and $\dot{M}_{\rm w}=2L_{\rm w,k}/v_{\rm w}^2$ is the mass injection rate of the wind, with $L_{\rm w,k}$ being the kinetic luminosity of the wind. We assume $L_{\rm w,k}$ to be 5\% of the bolometric luminosity of the AGN $L_b$ following WLI, keeping constant before the AGN quenches. 
Note that the sound speed in the shocked wind region is $\sim \sqrt{P_{\rm sw}/\rho_{\rm sw}}\simeq v_{\rm w}-v_{\rm sw}'$, which is generally larger than $v_s$. Thus, the sound-crossing time is shorter than the dynamical timescale and this validates the previous assumptions of a homogeneous density, temperature, and thermal pressure in this region. 

The total thermal energy in the shocked wind region or region~b is then $E_{\rm t,sw}=\frac{3}{2}P_{\rm sw}V_{\rm sw}=\frac{1}{2}M_{\rm sw}(v_{\rm w}-v_{\rm sw}')^2$ with $V_{\rm sw}=\frac{4}{3}\pi(R_s^3-R_{rs}^3)$ being the volume of the shocked wind. Similarly, we have the total thermal energy in the shocked gas (region c) is $E_{\rm t,sg}=\frac{1}{2}M_{\rm sg}v_s^2$. $M_{\rm sw}\simeq \dot{M}_{\rm w}(R_s/v_s-R_{\rm rs}/v_{\rm w})$ is the mass of the shocked wind at time $t\sim R_s/v_s$ and  $M_{\rm sg}=4\pi\int_0^{R_{s}} r^2\rho_{g}(r)dr$ is the total mass of the ambient gas swept-up by the forward shock, with $\rho_g$ being the density profile of the gas. WLI and WLII adopted a broken power-law distribution for $\rho_g$, i.e, $\rho_g \propto R^{-2}$ if $R<R_{\rm disk}$ and $\rho_g \propto R^{-3.95}$ if $R>R_{\rm disk}$ for a disk component and halo component of the gas respectively, where the disk radius is related to the virial radius of the galaxy by $R_{\rm disk}=0.04R_{\rm vir}$. However, such a setup will lead to an extremely high gas density at small radius. Usually, the average number density of ISM is approximately constant in the disk of a galaxy and not significantly larger than $\sim 10^4\,\rm cm^{-3}$ even for the starburst region of ULIRGs which are rich of gas (such as the core of Arp 220, \citealt{Downes98,Peng16}). Thus, we add a flat core in the density profile, i.e., $\rho_0=1000m_p (L_b/10^{45}\rm ergs^{-1})^{1/3}\, \rm cm^{-3}$, for the inner 100\,pc. On the other hand, there are intergalactic medium surrounding galaxies, so the gas density may not keep decreasing at large radius and we assume the gas density beyond the virial radius to be $\rho_g(R>R_{\rm vir})=\rho_g(R=R_{\rm vir})$. In summary, the adopted gas density profile in our calculation is
\begin{equation}\label{eq:density}
\rho_g=\left\{
\begin{array}{llll}
\frac{\rho_0}{1+\left(R/100\rm pc\right)^2}\, {\rm cm^{-3}}, ~~ R<R_{\rm disk},\\
\frac{\rho_0}{1+\left(R_{\rm disk}/100\rm pc\right)^2}\left(\frac{R}{R_{\rm disk}}\right)^{-3.95}\, {\rm cm^{-3}}, ~~R_{\rm disk}<R<R_{\rm vir}\\
\frac{\rho_0}{1+\left(R_{\rm disk}/100\rm pc\right)^2}\left(\frac{R_{\rm vir}}{R_{\rm disk}}\right)^{-3.95}\, {\rm cm^{-3}}, ~~ R>R_{\rm vir}
\end{array}
\right.
\end{equation}
The virial radius of a galaxy of a dark matter halo with mass $M_{h}$ can be given by 
\begin{eqnarray}
R_{\rm vir}&=&170\,h^{-2/3}\left[\left(\Omega_M(1+z)^3+\Omega_\Lambda\right)\Delta_{\rm c}/18\pi^2(1+z)^3\right]^{-1/3}\,\nonumber\\
&\times&\left(\frac{M_{\rm h}}{10^{12} M_\odot}\right)^{1/3}(1+z)^{-1} \,\rm ~kpc
\end{eqnarray}
where $\Delta _c$ is typically defined as the ratio of the average gas density within the virial radius of the galaxy to the critical density, and it is value is $\simeq 18\pi^2$ for a flat universe. 

We then can calculate the dynamic evolution of the shocked ambient gas, whose motion is governed by 
\begin{equation}\label{eq:mom}
\frac{d}{dt}\left(M_{\rm sg}v_{\rm sw}\right)=-\frac{GM_{G}(<R_{s})M_{\rm sg}}{R_{s}^2}+4\pi R_{s}^2\left(P_{\rm sw}-P_g \right)
\end{equation}
where the first term in the right-hand side of the equation considers the gravity on the expanding shell, exerted by the total gravitational mass within $R_s$, including the supermassive black hole (SMBH), the dark matter and the self-gravity of the shell of the swept-up gas itself, i.e., $M_G(<R_s)=M_{\rm BH}+M_{\rm DM}(<R_s)+M_{\rm sg}(<R_s)/2$. $M_{\rm DM}(<R_s)$ is calculated based on the NFW profile \citep{NFW96} and the total mass of the dark matter halo $M_h$. The value of $M_h$ and $M_{\rm BH}$ are related to the AGN's bolometric luminosity $L_b$, which are referred to the treatment by WLI (see also Appendix A.1 for the details). $P_g$ is the thermal pressure of the unshocked ambient gas, which tends to resist the expansion of the shell. The pressure is related to the gas density and temperature by $P_g=n_gkT_g$. Here, $T_g$ can be found by the hydrostatic equilibrium of the gas in the galaxy:
\begin{equation}
\frac{dT_g}{dR}=\frac{GM_{G}m_p}{kR^2}-\frac{T_g}{n_g}\frac{dn_g}{dR}.
\end{equation}
Note that at small radii the gases may not be in the hydrostatic equilibrium, since the density of the gas is so high that the cooling time is short, leading to a cooling inflow which feeds the activity of the SMBH and the star formation. So the temperature of the gases in the inner galaxy is probably lower than that estimated from the hydrostatic equilibrium. However, such an overestimation of the gas temperature will not have a significant effect on the evolution of the shocked gases, because the deceleration of the shocked gas in the inner galaxy is mainly caused by the increasing mass of the swept-up shell. On the other hand, the swept-up shell is pushed forward by the thermal pressure of the shocked wind $P_{\rm sw}$. { As we mentioned earlier,} it is related to the total thermal pressure in the shocked wind by
\begin{equation}\label{eq:PErelation}
E_{\rm t, sw}=\frac{3}{2}P_{\rm sw}V_{\rm sw}=2\pi P_{\rm sw}(R_s^3-R_{\rm sw}^3),
\end{equation}
while the changing rate of the thermal energy is generally determined by the energy injection from the wind, the work done on the swept-up shell and the radiative cooling of the shocked wind. {Considering the two-temperature effect, protons and electrons may have different temperatures and undergo different energy loss processes. Denote the thermal energies for protons and electrons by $E_{p,\rm sw}$ and $E_{e,\rm sw}$ respectively (with $E_{\rm t, sw}=E_{p,\rm sw}+E_{e,\rm sw}$), we have 
\begin{equation}\label{eq:th_rate_p}
\frac{dE_{p,\rm sw}}{dt}=\frac{1}{2}\dot{M}_{\rm w}(1-\frac{v_{\rm rs}}{v_{\rm w}})(v_{\rm w}^2-v_{\rm sw}'^2)-4\pi R_s^2P_{p,\rm sw}v_s-L_{\rm Cou}
\end{equation}
for protons. The first term in the right-hand side of the equation means the injection from the wind.  Note that the thermal energy injection rate is not the kinetic luminosity of the wind $\frac{1}{2}\dot{M}_{\rm w}v_{\rm w}^2$. This is because, firstly, the reverse shock also moves forward at a speed of $dR_{\rm rs}/dt=v_{\rm rs}$ and hence only a fraction of $1-v_{\rm rs}/v_{\rm w}$ of the wind material can inject into the shocked wind region in unit time, and secondly, a part of the energy, $\frac{1}{2}\dot{M}_{\rm w}v_{\rm sw}'^2$, goes into the kinetic energy of the shocked wind. The second term represents adiabatic losses due to the expansion or the work done on the swept-up shell. The third term $L_{\rm Cou}=(E_{p,\rm sw}-E_{e, \rm sw})/t_{pe}$ is the Coulomb cooling rate of protons due to collisions with electrons. $t_{pe}$ is the timescale for electrons and protons achieve equipartition. In principle, protons also suffer from the radiative cooling via Compton scattering, synchrotron radiation and free-free emission. Since the efficiencies of these cooling processes are very low for protons, we simply neglect them here. But this may not be true for electrons. Thus, we have 
\begin{eqnarray}\label{eq:th_rate_e}
\frac{dE_{e,\rm sw}}{dt}&=&\frac{1}{2}\dot{M}_{\rm w}\left(\frac{m_e}{m_p}\right)(1-\frac{v_{\rm rs}}{v_{\rm w}})(v_{\rm w}^2-v_{\rm sw}'^2)\nonumber\\
&-&4\pi R_s^2P_{e,\rm sw}v_s+L_{\rm Cou}-L_{C}-L_{\rm rad}.
\end{eqnarray}
The first two terms in the right-hand side have the same meaning with those in Eq.~(\ref{eq:th_rate_p}). The third term $L_{\rm Cou}$ is also exactly the same with that in Eq.~(\ref{eq:th_rate_p}) but with an opposite sign since the Coulomb collision between protons and electrons serve as the heating term for electrons. The fourth term $L_{\rm C}$ is the Compton cooling/heating rate through Compton scattering \citep{Sazonov01, FG12}. The last term $L_{\rm rad}=E_{e,\rm sw}(1/t_{\rm ff}+1/t_{\rm syn})$ is the radiative cooling rate due to free-free emission and synchrotron emission with $t_{\rm ff}$ and $t_{\rm syn}$ being the cooling timescales (see Appendix A.2 for details).}

Based on Eqs.~(\ref{eq:mom})--(\ref{eq:th_rate_e}), we obtain the evolution of various quantities after we find out the initial condition of those quantities (see Appendix A.3 for details). Fig.~\ref{fig:evol} shows the evolution of forward shock speed $v_s$, the thermal pressure in the shocked wind region $P_{\rm sw}$ and the energy injection rate of thermalized protons from the forward shock $L_{\rm th}$ as well as the density profile of ambient gas for reference, for AGNs with $L_b=10^{42}\,$erg/s, $10^{45}\,$erg/s, and $10^{48}\,$erg/s at $z=1$. In panel a, dashed curves show the case of a constant wind injection and we can see three breaks in the curves for $v_s$, because of the change in the profile of the gas density. A self-similar analytical solution to $v_s$, assuming that half kinetic energy injected by the wind constantly goes into kinetic motion of the swept-up shell, reads $v_s\propto R^{(\alpha-2)/3}$ \citep{FG12} with $\alpha$ the power-law index of the density profile. For the employed profile, we have $\alpha=0,2,3.95(\simeq 4),0$, leading to $v_s\propto R^{-2/3}, R^0, R^{2/3}, R^{-2/3}$, respectively, for the core region, the disk, the halo, and outside virial radius (dashed curves). 
Our results do not deviate much from this analytical one, implying radiation losses are not severe and the flow is ``energy-driven''. The dotted vertical lines in the figures mark the Salpeter time $t_{\rm sal}\simeq 4\times 10^7\,$yr for a radiative efficiency of 0.1, which is regarded as the lifetime of the quasars \citep[e.g.,][]{Yu02}. Beyond the lifetime, we assume the AGN shuts off and hence the injection of the wind vanishes, as in WLI. We can see the swept-up shell of shocked gas will not stop immediately, and continue to expand but with a different dynamic evolution quickly after the Salpeter time (solid curves). Without the further energy injection into the region of shock wind, the thermal energy therein depletes quite fast, as can be seen in panel~b, and the forward shock starts to be decelerated after losing the push by the thermal pressure of the shocked wind. 
Note that given the temperature of gas in the halo of a galaxy to be $\sim 10^{6}$K, the sound speed is supposed to be a few times $10^6\,$cm/s. Thus, the forward shock may disappear at late times for low-luminosity AGNs (e.g., $L_b=10^{42}$erg/s). Later, for the calculation of gamma-ray and neutrino production, we will use the evolution with the quenching of AGN considered.  

\begin{figure*}[htbp]
\centering
\includegraphics[width=0.9\textwidth]{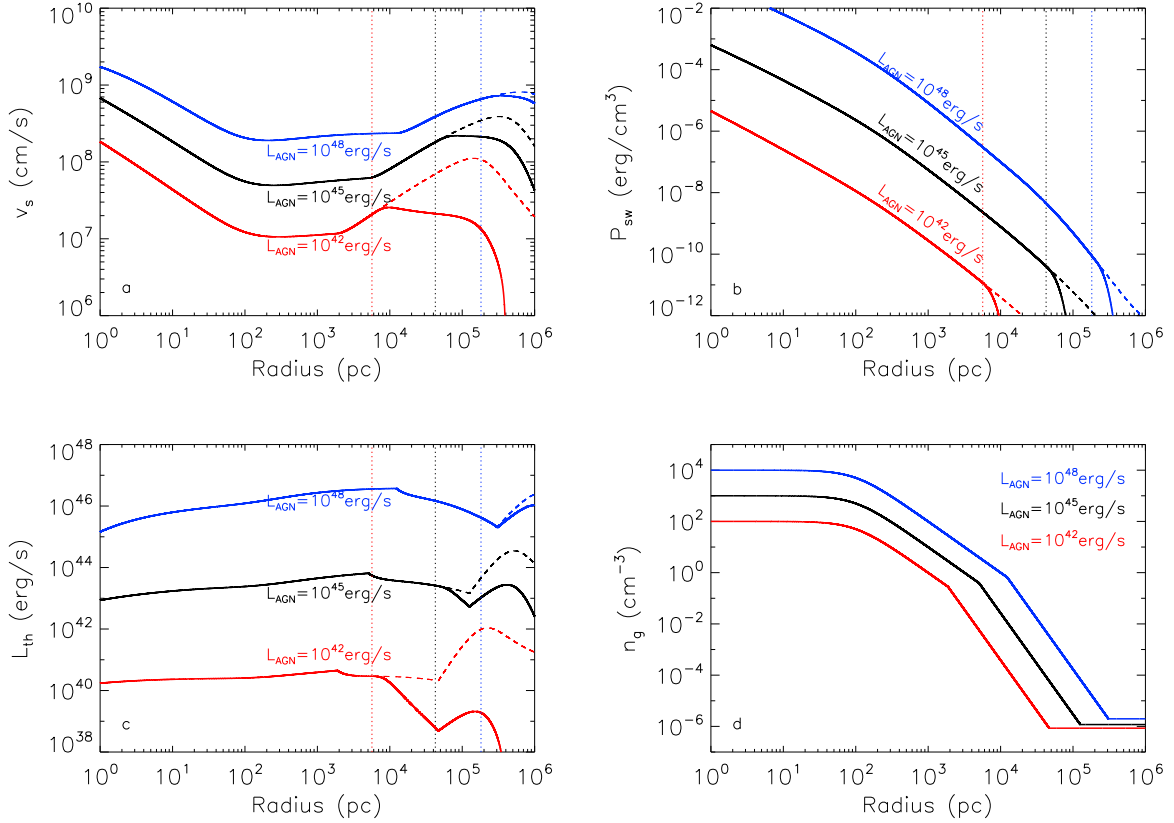}
\caption{Evolution of forward-shock velocity ($v_s$, panel a), total thermal pressure of the shocked wind ($P_{\rm sw}$, panel b), energy injection rate of thermal protons by the forward shock ($L_{\rm th}$, panel c), and the density profile of the ambient medium (panel d) for AGNs at $z=1$ with $L_b=10^{42}$\,erg/s (red), $10^{45}\,$erg/s (black), $10^{48}\,$erg/s (blue), respectively. The dashed curves are for the evolution without the AGN shutoff. The red, black, and blue vertical dotted lines represent the shock radii after propagating a time of $4\times 10^7\,$yr or the Salpeter time in the case of $L_b=10^{42}\,$erg/s, $L_b=10^{45}\,$erg/s, and $L_b=10^{48}\,$erg/s respectively.}\label{fig:evol}
\end{figure*}

The energy injection rate of thermalized protons from the forward shock is given by $L_{\rm th}=4\pi R_s^2 u_{\rm sg,t}v_s$ where $u_{\rm sg,t}=\frac{9}{32}\rho_{\rm sg}v_s^2$ is the thermal energy density of the shock-compressed gas and $\rho_{\rm sg}=4\rho_g$. We assume a fraction $\epsilon_{\rm nt}=0.1$ of the thermal energy  injected per unit time is converted to non-thermal energy of relativistic protons via the forward shock, i.e., the cosmic ray luminosity $L_{\rm CR}=\epsilon_{\rm nt}L_{\rm th}$. If we use the analytical solution of the dynamic evolution for  estimation, we have $L_{\rm CR}\propto L_{\rm th}\propto R^0$, and this behavior is more or less consistent with our results as shown in panel~c.

\begin{figure*}[htbp]
\centering
\includegraphics[width=0.9\textwidth]{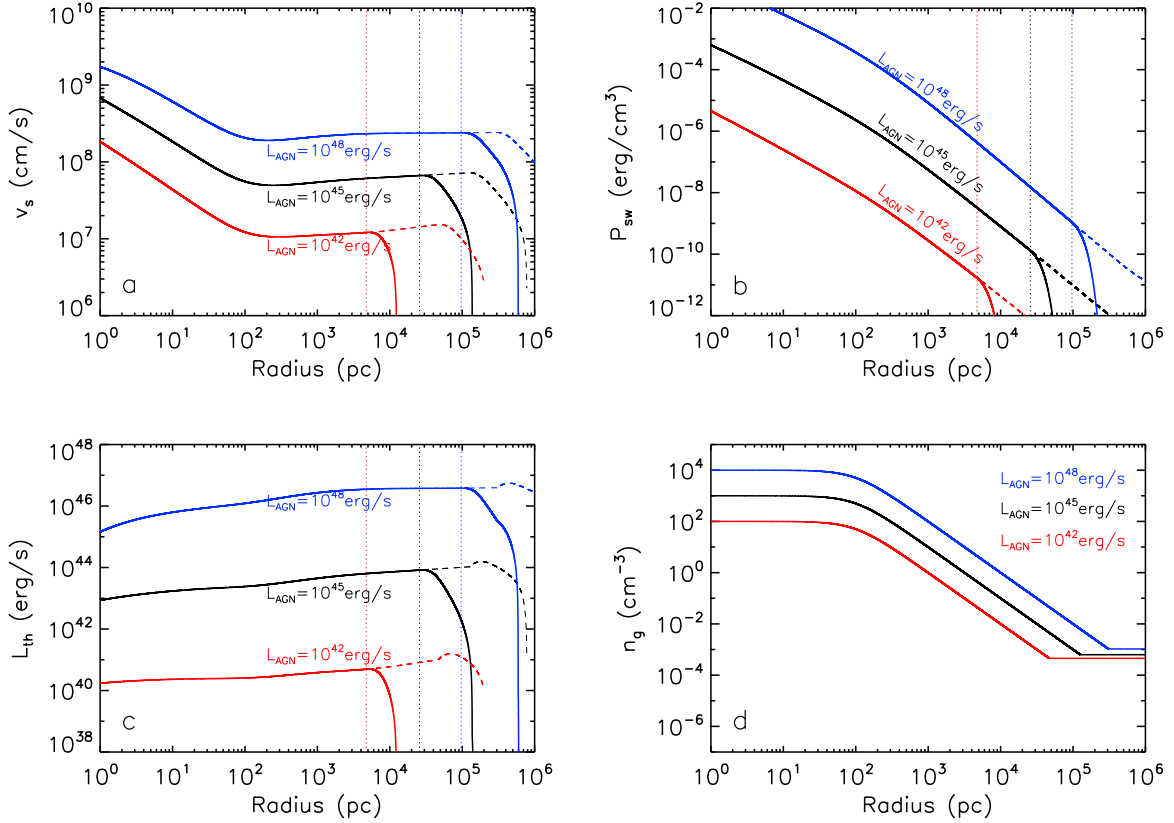}
\caption{Same as Fig.~\ref{fig:evol}, but for the gas density profile without a further break in halo (Eq.~\ref{eq:density_ist}), as shown in panel d.}\label{fig:evol_ist}
\end{figure*}

Given the adopted gas density profile, the total gas mass in the host galaxy is about $5-10\%$ of the total  mass (including dark matter), while the cosmic mean baryon fraction is about 16\%. This may not be unreasonable given the ``missing'' baryon problem found in many galaxies (e.g., see \citealt{Bregman07} and reference therein). { Actually, although our assumption on the gas density is more conservative than those in the previous literatures (WL1, WL2;\citealt{Lamastra17}), our employed gas density profile might still be an overestimation of gas content for some AGNs' host galaxies. From Eq.~(\ref{eq:density}), we can see that the column density ranges from $N_H=6\times 10^{22}\,\rm cm^{-2}$ for the AGN with the lowest luminosity in our calculation (i.e., $L_b=10^{42}\,$erg/s) to $N_H=6\times 10^{24}\,\rm cm^{-2}$ for the most luminous one ($L_b=10^{48}\,$erg/s). However, observationally, a large fraction of AGNs (mainly Seyferts) have a smaller column density than $N_H=6\times 10^{22}\,\rm cm^{-2}$, at any redshift or luminosity \citep[e.g.][]{Ueda03, LaFranca05, Tozzi06, Lusso12, Ueda14}. Although there also exists many AGNs with column density larger than $6\times 10^{24}\, \rm cm^{-2}$, such high column densities are found to be predominantly caused by the parsec-scale dusty torus \citep[e.g.][]{Fukazawa11, Goulding12, Buchner17, LiuT17}, depending on the covering factor and viewing angles. Note that the important quantity relevant to the gamma-ray and neutrino production is the gas content associated with the galactic disk and halo rather than the dusty torus. Thus, the adopted gas density profile may lead to a significant overestimation of the gas column density and the subsequent gamma-ray and neutrino production, although such a high-density environment would be reasonable for AGNs coexisting with starbursts \citep{Tamborra14}.} 
On the other hand, if one wants to keep the gas fraction similar to the cosmic mean value, one may take a larger value for $\rho_0$ or assume a shallower decrease of the density profile in the halo. As a reference, we here also consider the dynamical evolution of the wind bubble in an alternative case in which there is no further break in the density profile of gas in the halo, i.e., 
\begin{equation}\label{eq:density_ist}
\rho_g=\left\{
\begin{array}{llll}
\frac{\rho_0}{1+\left(R/100\rm pc\right)^2}\, {\rm cm^{-3}}, ~~ R<R_{\rm vir},\\
\frac{\rho_0}{1+\left(R_{\rm vir}/100\rm pc\right)^2}\, {\rm cm^{-3}}, ~~ R>R_{\rm vir}
\end{array}
\right.
\end{equation}
The results are shown in Fig.~\ref{fig:evol_ist}. In this case, the forward shock will undergo a constant deceleration. Note that in this case the gas mass fraction can even reach $\sim 30-60\%$. Given such an abundant gas content, the effective $pp$ reaction optical depth increases at large radius. It is certainly not a realistic case, but we may regard it as an upper limit, and we can see later in Fig.~\ref{fig:lc} that the resultant gamma-ray or neutrino flux does not increase much compared to the case shown in Fig.~\ref{fig:evol}.

\section{Gamma-ray and Neutrino Production}
The spectrum of produced gamma-ray and neutrinos depends on the parent protons. In this work, we consider two different forms of the proton spectrum. Firstly, following WLI and WLII, the differential cosmic-ray proton density injected at a radius $R$ is assumed to be a single power-law (SPL) with a cutoff at the highest energy, i.e.
\begin{equation}\label{eq:spl}
\begin{split}
N_p^{\rm inj}(E_p,R)&\equiv\frac{dN}{dE_p}\\
&=N_0(R)\left(\frac{E_p}{100\,\rm TeV}\right)^{-\Gamma_{\rm CR}}{\rm exp}\left(-E_p/E_{p, \rm max} \right)
\end{split}
\end{equation}
where $\Gamma_{\rm CR}$ is the power-law index for cosmic-rays, and $E_{p, \rm max}$ is the maximum achievable proton energy in the forward shock. 

\begin{figure}[htbp]
\centering
\includegraphics[width=0.5\textwidth]{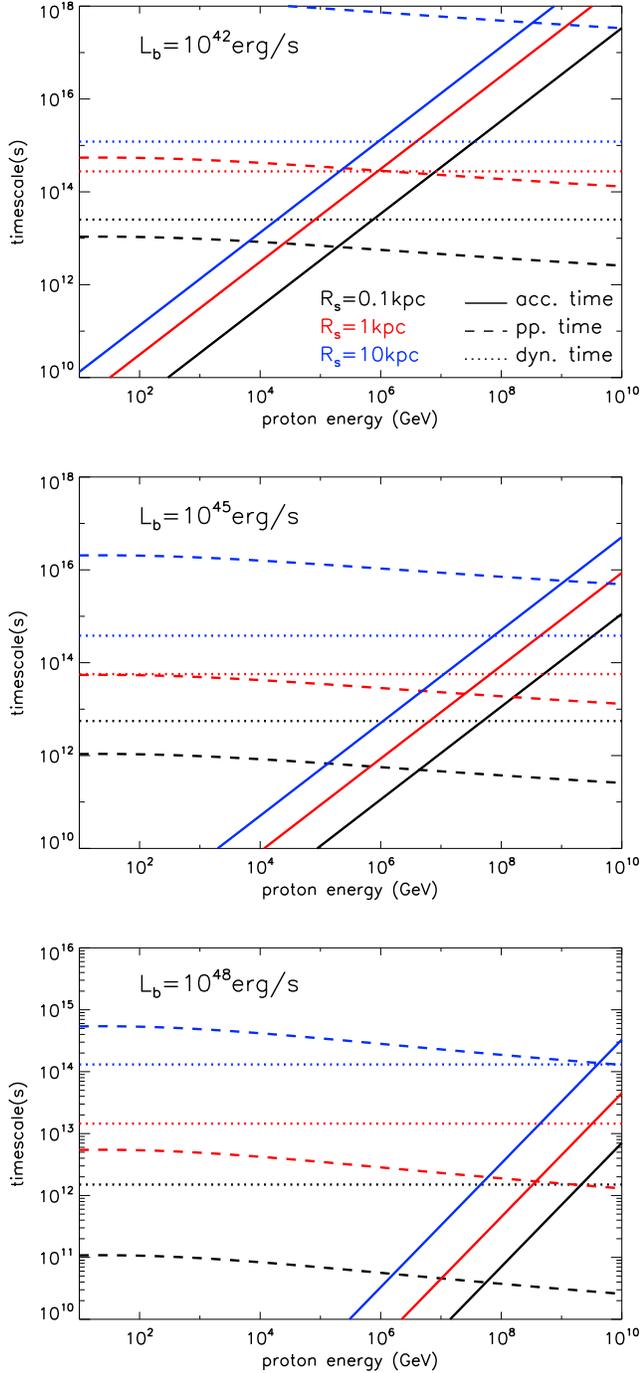}
\caption{Timescales of different processes for AGNs with bolometric luminosities $L_b=10^{42}\,$erg/s (top panel), $L_b=10^{45}\,$erg/s (middle panel), and $L_b=10^{48}\,$erg/s (bottom panel). Solid, dashed, and dotted lines represent acceleration, $pp$ reaction, and dynamical timescales, respectively. Black, red, and blue lines are for different shock radii, as marked inside the top panel.}\label{fig:timescale}
\end{figure}

As another form of the proton spectrum, we introduce a break below $100\,$TeV where the spectrum becomes flat, i.e., a broken power-law (BPL) with a cutoff,
\begin{equation}\label{eq:bpl}
N_p^{\rm inj}(E_p,R)=N_0(R)\left\{
\begin{array}{ll}
\left(\frac{E_p}{100\,\rm TeV}\right)^{-2},~~~~ E_p<100\,{\rm TeV}\\
\begin{split}
\left(\frac{E_p}{100\,\rm TeV}\right)^{-\Gamma_{\rm CR}}&{\rm exp}\left(-E_p/E_{p, \rm max} \right)\\
&, E_p>100\,\rm TeV.
\end{split}
\end{array}
\right.
\end{equation}
{The BPL spectrum can keep the secondary neutrino spectral index consistent with the observed one \citep{IC15} while in the meantime reduces the amount of cosmic rays at low energy and, consequently, the gamma-ray production below 10\,TeV, compared to the SPL spectrum with the same $\Gamma_{\rm CR}$.} $E_{\rm max}$ can be found by equating the shock acceleration timescale $t_{\rm acc}=20E_p c/3eBv_s^2$ and the minimum between the dynamical timescale $t_{\rm dyn}\approx R/v_s$ and the $pp$ cooling time $t_{pp}=1/(0.5\sigma_{pp}n_{\rm sg}c)$ where $\sigma_{pp}$ is the cross section for $pp$ interactions and $n_{\rm sg}(R)=4n_g(R)$ is the number density of the compressed gas by the forward shock. The magnetic field $B=[12\pi \epsilon_B n_{\rm sg}(R)kT_{\rm sg}(R)]^{1/2}$ where $\epsilon_B$ is the equipartition parameter for the magnetic field which is fixed at 0.1 at any $R$, $k$ is the Boltzmann constant, and $T_{\rm sg}(R)=3 m_p v_s(R)^2 /16k$ is the temperature of the shocked ambient gas. We have $E_{p,\rm max}\simeq 2\times 10^{17}(\frac{\epsilon_B}{0.1})^{1/2}(\frac{n_g}{1000\,\rm cm^{-3}})^{1/2}(\frac{R_s}{100\,\rm pc})(\frac{v_s}{10^8\,\rm cm~s^{-1}})^2\,$eV if the dynamical time is shorter and $E_{p,\rm max}\simeq 10^{17}(\frac{\epsilon_B}{0.1})^{1/2}(\frac{n_g}{1000\,\rm cm^{-3}})^{-1/2}(\frac{\sigma_{pp}}{50\,\rm mb})^{-1}(\frac{v_s}{10^8\,\rm cm~s^{-1}})^3\,$eV if the $pp$ cooling time is shorter. One can find the timescale of relevant processes in Fig.~\ref{fig:timescale} for AGNs with different bolometric luminosities. 
Note that there should be a pre-factor $>1$ in the expression for the acceleration timescale since the spatial diffusion of accelerated particles can be far from the Bohm limit. This can lead to the maximum proton energy easily dropping below $10-100$\,PeV and hence creating difficulties in the explanation of PeV neutrinos, although it does not have a significant influence on the production of GeV photons. To compare our results with those in previous literature, we will consider the maximum proton energy obtained in the Bohm limit in the following calculation. 
The normalization factor $N_0(R)$ is found by assuming that a fraction $\epsilon_{\rm nt}=0.1$ of the newly injected thermal energy in the shocked gas goes into the nonthermal energy of accelerated protons, i.e., $L_{\rm CR}=\epsilon_{\rm nt}L_{\rm th}$ as we mentioned in the previous section. Note that $L_{\rm CR}=4\pi R^2v_su_{{\rm CR},p}^{\rm inj}$ with $u_{{\rm CR},p}^{\rm inj}\equiv \int E_pN_p^{\rm inj}(E_p)dE_p$ and $L_{\rm th}=4\pi R^2v_su_{\rm sg, t}$ with $u_{\rm sg,t}=\frac{3}{2}n_{\rm sg}(R)kT_{\rm sg}(R)$. So we have $\int E_pN_p^{\rm inj}(E_p)dE_p=\frac{3}{2}\epsilon_{\rm nt}n_{\rm sg}(R)kT_{\rm sg}(R)$.

We assume that most accelerated protons are well confined in the downstream of the forward shock (or the region b) and interact with gas therein during the advection, since the Larmor radius of a proton reads 
\begin{eqnarray}
r_L\simeq && 10^{15}\left(\frac{E_p}{10^{15}\, \rm eV}\right)\left(\frac{\epsilon_B}{0.1}\right)^{-1/2}\left(\frac{n_g}{1000\,\rm cm^{-3}}\right)^{-1/2}\nonumber\\ 
&\times& \left(\frac{v_s}{10^8\,\rm cm~s^{-1}}\right)^{-1}\, \rm cm 
\end{eqnarray}
which is much smaller than the shock radius. The spectra of secondaries in $pp$ collisions are calculated based on the semianalytical method developed by \citet{Kelner06}. In the optically thin limit, the spectral index of gamma rays and neutrinos, $\Gamma$, is roughly estimated to be $\Gamma\approx\Gamma_{\rm CR}-0.1$. However, in the calorimetric limit, we expect $\Gamma\approx\Gamma_{\rm CR}$.  
For the clarity of the expression, let's define an operator ${\mathcal F}_\gamma$ that can obtain the gamma-ray emissivity by $N_\gamma(E_\gamma)={\mathcal F}_{\gamma}\{N_p(E_p),n_{\rm sg}\}$ (basically it is the same as for neutrino emissivity; we just need to use another operator ${\mathcal F}_\nu$), with
\begin{equation}
{\mathcal F}_{i}(E_i)= cn_{\rm sg}\int_{E_i}^{\infty}\sigma_{pp}{dN_p}{dE_p}(E_p)F_i(\frac{E_i}{E_p},E_p)\frac{dE_p}{E_p}
\end{equation}
where $i$ could be $\gamma$ or $\nu$, and $F_i$ is the spectrum of the secondary $\gamma$ or $\nu$ in a single collision. This presentation works for $E_p\gtrsim 100\,$GeV, while for $E_p < 100\,$GeV a $\delta$-functional approximation for the energy of produced pions can be used to obtain the secondary spectrum \citep{Kelner06}, i.e.
\begin{equation}
\begin{split}
{\mathcal F}_{i}(E_i)&=2cn_{\rm sg}\frac{\tilde{n}}{K_\pi}\int_{E_{i, \rm min}}^{\infty}\sigma_{pp}\left(m_p+\frac{E_\pi}{K_\pi}\right)\\
&\times \frac{dN_p}{dE_p}\left(m_p+\frac{E_\pi}{K_\pi}\right)\frac{dE_\pi}{\sqrt{E_\pi^2-m_\pi^2}}
\end{split}
\end{equation}
where $E_\pi$ is the energy of pions and the pion rest mass $m_\pi \simeq 135\,$MeV for gamma-ray production and $m_\pi\simeq 140\,$MeV for neutrino production. $E_{i,\rm min}=E_i/\zeta_i+\zeta_i m_\pi^2/4E_i$, with $\zeta_\gamma=1$ and $\zeta_\nu=1-m_\mu^2/m_\pi^2=0.427$ ($m_\mu\simeq 106\,$MeV is the muon rest mass), $K_\pi=0.17$, and $\tilde{n}$ is a free parameter that is determined by the continuity of the flux of the secondary particle at $100\,$GeV. To get the differential isotropic gamma-ray luminosity for the shock front at $R$, we need to integrate over the emission of all the protons injected in history, i.e.,
\begin{equation}
L_\gamma(E_\gamma,R)=E_\gamma^2\int_0^R {\mathcal F}_\gamma\{N_p(E_p,r;R),n_{\rm sg}(R)\}4\pi r^2dr
\end{equation}
where $N_p(E_p,r;R)$ represents the differential number density of protons which were injected at a radius $r(<R)$ when the shock front is at $R$. Note that $N_p(E_p,r;R)$ is different from $N_p^{\rm inj}(E_p,R)$ because of energy losses due to $pp$ interactions and adiabatic losses that will turn out to be important. 
After some approximations (see Appendix B for details), we arrive at
\begin{eqnarray}\label{eq:Np}
N_p(E_p,r;R)&=&N_p^{\rm inj}(E_p,r){\rm exp}[-(1-2^{1-\Gamma_{\rm CR}})\tau_{pp}(E_p,r,R)\nonumber\\
&-&(\Gamma_{\rm CR}-1)\tau_{\rm ad}(r, R)
\end{eqnarray}
where $\tau_{pp}(E_p,r,R)=\int_t(r)^t(R) \sigma_{pp}(E_p)n_{\rm sg}(r'(t))cdt$ and $\tau_{\rm ad}(r,R)=\int_t(r)^t(R) v_s(t)dt/r'(t)$. We note that photopion production interactions between accelerated protons and radiation fields of AGNs could be important at small radii, especially in the presence of AGN jets. 
However, we neglect this process because the number of accelerated protons at small radii is quite limited and also because this phase will not last long since the shock speed is high. 

\begin{figure}[htbp]
\centering
\includegraphics[width=1\columnwidth]{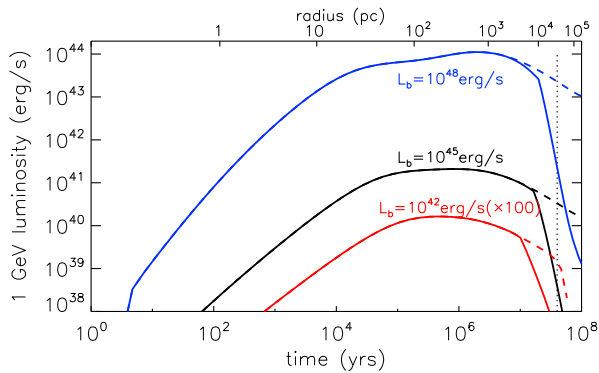}
\caption{The light curve of 1\,GeV photons produced in an AGN-driven wind with the AGN's bolometric luminosity being $10^{45}\,$erg/s (black), $10^{42}\,$erg/s (red), $10^{48}\,$erg/s (blue), located at $z=1$. The solid curves show the light curves corresponding to the results with the gas density profile and evolution presented in Fig.~\ref{fig:evol}, while dashed curves correspond to those in Fig.~\ref{fig:evol_ist}. The vertical dotted line shows the Salpeter time $t_{sal}=4\times 10^7$yr. The top x-axis marks the corresponding shock radius at certain time for $L_b=10^{45}\,$erg/s.}\label{fig:lc}
\end{figure}

Then, the gamma-ray light curve of an AGN wind can be calculated based on above equations and the results are shown in Fig.~\ref{fig:lc}. The solid curves represent the 1\,GeV gamma-ray light curve produced by an AGN wind under the evolution and gas density profile presented in Fig.~\ref{fig:evol}. The top x-axis marks the corresponding shock radius at certain time for $L_b=10^{45}\,$erg/s. The gamma-ray luminosities reach the maximum when the shocks propagate to the radius of $\sim 0.1-1$\,kpc. The light curve also shows a plateau-like behavior in this range because the swept-up shell is a proton calorimeter while the cosmic-ray luminosity is more or less constant in this range. The wind is not well decelerated at smaller radii, while the gas density becomes very small at larger radii. As a result, the neutrino and gamma-ray luminosities are relatively low at these locations (see Fig.~\ref{fig:timescale} for reference).  We find most energies ($\sim Lt$) are emitted when the shock is around $\lesssim 10\,$kpc. The average luminosity within $t_{\rm sal}$ is about 5 times smaller than the peak luminosity for $L_b=10^{42}\,$ergs/, 3 times smaller for $L_b=10^{45}\,$erg/s and for $L_b=10^{48}\,$erg/s. For reference, the dashed curves show the results in the case that no break appears in the gas density profile in the halo, i.e., corresponding to the case presented in Fig.~\ref{fig:evol_ist}. The average luminosity in this unrealistic case increases only by a factor of $\sim 2$. We note that the neutrino light curve should follow the same temporal behavior as that of the gamma rays.

{Given the setup in this work, the AGN redshift ($z$) only influences the virial radius $R_{\rm vir}$ and the correlated disk radius $R_{\rm disk}=0.04R_{\rm vir}$. For the same AGN luminosity, a larger redshift leads to a smaller $R_{\rm vir}$ and $R_{\rm disk}$. As a result, the total mass content is reduced in the halo while the gas distribution is still the same in the disk. Therefore, a larger redshift leads to a less efficient gamma-ray/neutrino production in the halo. From the perspective of the lightcurve, the position of the decline in the lightcurve at $\lesssim 10\,$kpc should appear earlier for larger $z$ and vice versa. In reality, the density may also positively scale as redshift and results in a larger gamma-ray/neutrino production for higher redshift AGN host galaxies. In principle, a more careful treatment is necessary, such as done in \citet{Yuan17}. }

We are aware of that after an AGN shuts off, the forward shock may still expand into the ambient gas and accelerate protons. However, the host galaxy would no longer be regarded as a quasar-type or Seyfert-type AGN for the current observers, although it may be left as a low-luminosity AGN with powerful jets. 
Since we are only concerned with the gamma-ray and neutrino fluxes from AGNs, we do not consider the production beyond $t_{\rm sal}$. On the other hand, even if we assume all the inactive galaxies were AGNs, their contribution to the diffuse gamma-ray and neutrino fluxes should be minor compared to that from AGNs at the present time. This is because that the AGN fraction is about $\sim 1\%$ among all the galaxies \citep{Haggard10}, while the emissivity of gamma rays or neutrinos from an inactive galaxy is far smaller than $1\%$ of the average emissivity during its active period.

\section{Contribution to Diffuse Neutrino and Gamma-ray Backgrounds}
In the previous section, we have examined the gamma-ray and neutrino light curves from a single AGN embedded in a dense ISM surrounded by a less dense halo. To obtain the diffusive gamma-ray/neutrino flux from AGNs throughout the universe, we need to sum up the contribution from AGNs with different luminosities and redshifts. Note that those AGN-driven wind bubbles should be at different stages of the evolution, so we need to take the average luminosity during their lifetime, which can be given by $\bar{L}_{\gamma/\nu}(E)=\int_0^{t_{\rm Sal}} L_{\gamma/\nu}[E,R(t)]dt/t_{\rm Sal}$. Finally, we have the diffuse gamma-ray flux 
\begin{eqnarray}
E_\gamma^2\Phi_\gamma(E_\gamma)&=&\frac{c}{4\pi H_0}\int\int \Psi(L_b,z)\frac{\bar{L}_{\gamma,L_b,z}\left[(1+z)E_\gamma ,L_b\right]}
{(1+z)^2E(z)}\nonumber\\
&\times&{\rm exp}\left[-\tau_{\gamma\gamma} (E_\gamma ,z)\right]dzdL
\end{eqnarray} 
where $E(z)=\sqrt{(1+z)^3\Omega_M+\Omega_\Lambda}$ and $\Psi(L_b,z)$ is given by \citet{Hopkins07}
\begin{equation}
\Psi(L_b,z)\equiv\frac{d\Psi}{d{\rm log}L_b}=\frac{\Psi_\star}{(L_b/L_\star)^{\gamma_1}+(L_b/L_\star)^{\gamma_2}},
\end{equation}
accounting for the number of AGNs per logarithmic luminosity interval per volume. We adopt the pure luminosity evolution model and the parameters are given by $\log\left(\Phi_\star/{\rm Mpc^3}\right)=-4.733$, $\log L_\star=\log L_0+k_{L,1}\xi + k_{L,2}\xi^2+k_{L,3}\xi^3$, $\xi=\log\left[(1+z)/(1+z_{\rm ref})\right]$, $z_{\rm ref}=2$, $\log (L_0/L_\odot)=12.965$, $k_{L,1}=0.749$, $k_{L,2}=-8.03$, $k_{L,3}=4.40$, $\gamma_1=0.517$, and $\gamma_2=0.296$. $\tau_{\gamma\gamma}(E_\gamma,z)$ is the gamma-ray opacity due to absorption by cosmic microwave background (CMB) and extragalactic background light (EBL) for a photon originated from redshift $z$ with a red-shifted energy $E_\gamma$ at the earth. We adopt an EBL model of moderate intensity provided by \citet{Finke10}. In WLI and WLII, they adopted the EBL model of \citet{Stecker06} which was already ruled out by the gamma-ray observations by Fermi-LAT and observations with imaging atmospheric Cherenkov telescopes \citep[e.g.][]{Fermi10_EBL, Orr11}. But to compare with their results, we also adopt this EBL model in our calculation for reference. Note that one should remove this term when calculating the diffuse neutrino flux.

\begin{figure*}[htbp]
\centering
\includegraphics[width=0.7\textwidth]{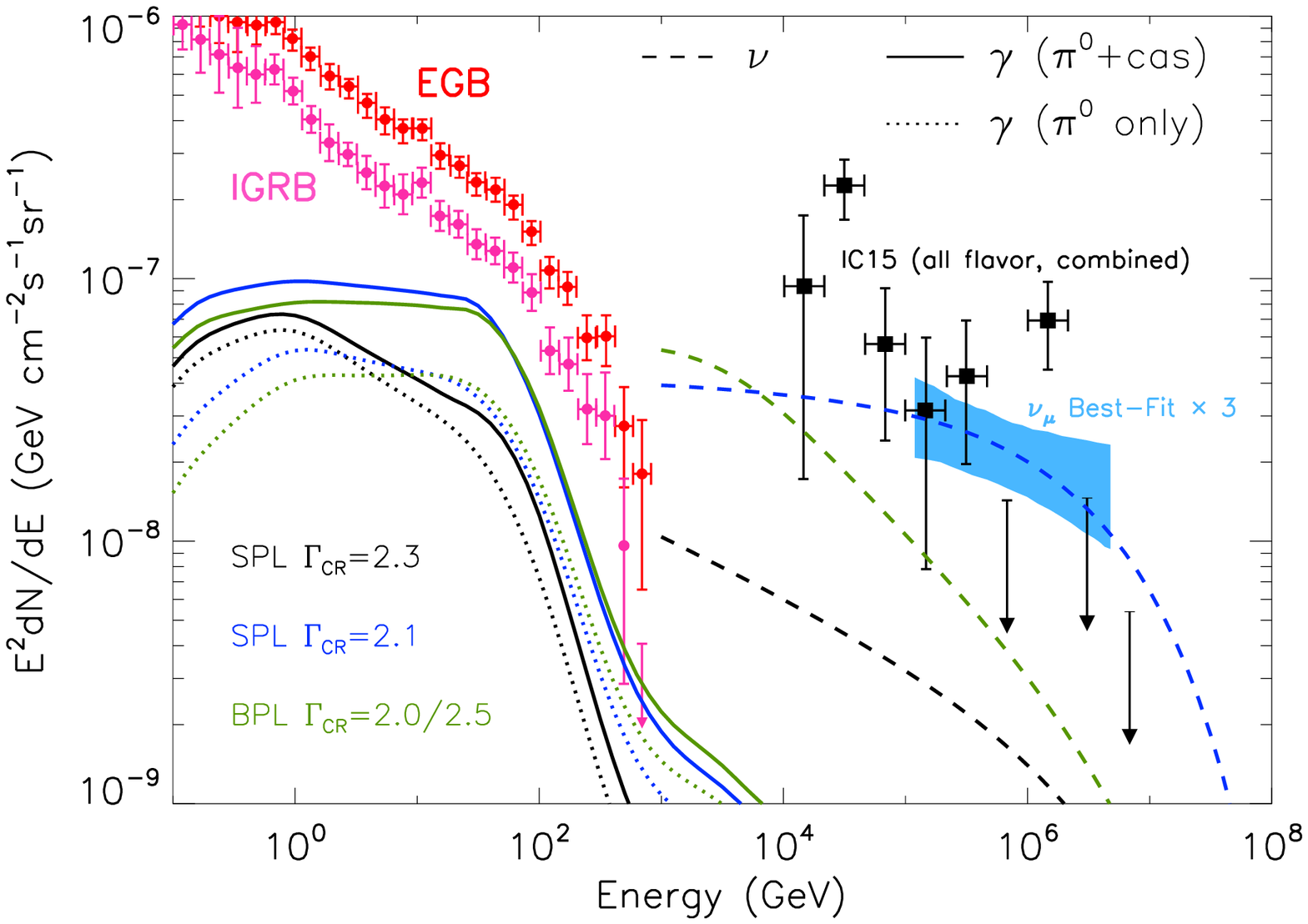}
\caption{Diffuse gamma-ray flux (solid curves) and all-flavor neutrino flux (dashed curves). The gamma-ray flux from direct $\pi^0$ decay is also shown (dotted curves). Different colors represent the cases for different proton spectral indices. The EBL model by \citet{Finke10} is adopted. The red and pink filled circles represent the Fermi-LAT EGB and IGRB data for foreground model A, respectively \citep{Fermi15}. 
The black filled squares are the astrophysical neutrino flux measured by IceCube \citep{IC15}, obtained from a combined maximum likelihood analysis while the blue shaded region corresponds to the 68\% C.L. allowed region for the muon (including anti-muon) neutrino flux with a single power-law model (\citealt{IC17_ICRC}, the original data have been multiplied by 3 to convert to an all-flavor flux, assuming a flavor ratio of 1:1:1).}\label{fig:flux_result}
\end{figure*}

After integrating over the luminosity in the range of $10^{42}-10^{48}\,$erg/s and redshift in the range of $z=0-5$, we can obtain the diffusive gamma-ray and neutrino backgrounds. Fig.~\ref{fig:flux_result} shows the results with different proton spectrum at injection, i.e., single power-law (SPL) spectrum with $\Gamma_{\rm CR}=2.3$ and $\Gamma_{\rm CR}=2.1$, and broken power-law (BPL) spectrum with $\Gamma_{\rm CR}=2$ below 100\,TeV and $\Gamma_{\rm CR}=2.5$ above 100\,TeV. No internal absorption of high-energy photons are considered, but electromagnetic cascades initiated by high-energy photons during the propagation in the intergalactic space are taken into account based on the EBL model of \citet{Finke10}. In this work, the calculation of electromagnetic cascades follows the simplified method described in \citet{Liu16}, and a sufficiently weak intergalactic magnetic field ($\lesssim1$~nG) is assumed so that cascades in the considered energy range will not be affected by synchrotron losses \citep[see][]{Murase12}. Given the total cosmic-ray luminosity, the GeV gamma-ray flux from direct $\pi^0$ decay in the case of $\Gamma_{\rm CR}=2.3$ is higher than those in the cases of $\Gamma_{\rm CR}=2.1$ and the BPL case. However, due to the contribution of the cascade emission whose energy production rate is the $\gtrsim100$GeV gamma-ray photons, the total GeV gamma-ray flux for $\Gamma_{\rm CR}=2.3$ becomes smaller than the latter two cases. 

Neutrinos are not affected during their propagation, except for adiabatic losses due to the expansion of the universe. Thus, if one extrapolates neutrino flux to the GeV range, the flux level is consistent with the $\pi^0$ gamma-ray flux.  
We can see that, in all three cases, the gamma-ray fluxes are significantly lower than the observed EGB flux at $1-10$\,GeV, constituting at most a fraction of $\lesssim 30\%$ of the EGB around 50\,GeV. On the other hand, the neutrino fluxes are lower than the best-fit IceCube flux at 10\,TeV by a factor of $5-20$. 
However, we note that in the case of a hard index, $\Gamma_{\rm CR}=2.1$, although the neutrino flux is about 5 times lower than the best-fit value at 10\,TeV, the flux above 100\,TeV is consistent with that inferred from through-going muon neutrino detection (assuming the neutrino flavor ratio to be 1:1:1). 
Indeed, the two-component scenario is possible, in which a hard component above $100\,$TeV \citep{IC15,IC16_mu} can be explained by cosmic-ray reservoir models, which may be even related to the sources of ultra-high-energy cosmic rays~\citep{Liu14, Murase16_PRD,Fang17}. 

Given the adopted luminosity function, we find that AGNs with luminosity around $10^{45}\,$erg/s make the most important contribution. This explains the softening of the neutrino spectrum at $\sim 1\,$PeV since the maximum proton energy is around 100\,PeV for $L=10^{45}\,$erg/s AGN when the shock is around 10\,kpc (see Fig.~\ref{fig:timescale}), where most energies are released as we discussed above. 
We reiterate that acceleration of $\sim 1-100\,$PeV protons is required to produce $10\,$TeV--$1\,$PeV neutrinos via inelastic $pp$ collisions. According to Fig.~\ref{fig:timescale}, protons with such high energies may not be achieved in the forward shocks of some AGN winds, especially when considering the Bohm diffusion fails to establish. A more realistic diffusion model would result in a much smaller maximum proton energy, probably leading to a cutoff in the produced neutrino spectrum below 10\,TeV. 

\subsection{Comparison to previous works}
\begin{figure*}[htbp]
\centering
\includegraphics[width=0.7\textwidth]{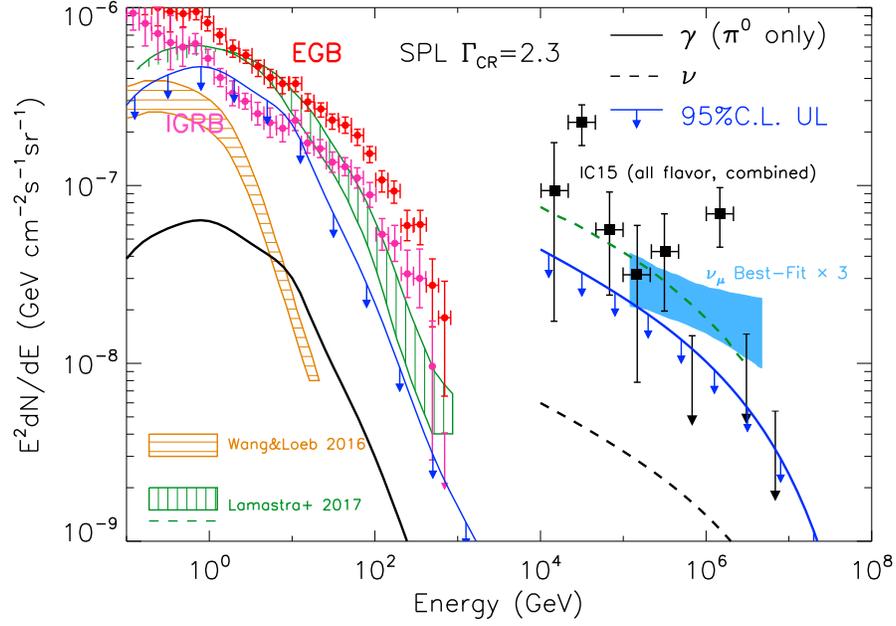}
\caption{Comparison of our result to the results in previous literature. An SPL with index of $\Gamma_{\rm CR}=2.3$ and the EBL model provided by \citet{Stecker06} are employed. The gamma-ray contribution from electromagnetic cascades is not considered. 
The blue curves with downward-pointing arrows present the upper limits for gamma-ray and neutrino fluxes when the 95\% C.L of the IGRB data is not violated. The yellow hatched region (dashed curve) is the gamma-ray (neutrino) flux obtained by WLII while the green hatched region (dashed curve) is the gamma-ray (neutrino) flux obtained by \citet{Lamastra17}.}\label{fig:flux_compare}
\end{figure*}

To compare our results with those in previous literature, we consider a case with the proton spectral index of $\Gamma_{\rm CR}=2.3$, with counting only the gamma-ray flux from $\pi^0$ decay, and employ the EBL model given by \citet{Stecker06}, which are adopted in WLI and WLII. The result is shown in Fig.~\ref{fig:flux_compare}. Our gamma-ray flux is several times lower than that obtained in WLII. 
This would be partly because they extrapolated a $R^{-2}$ profile for the gas density down to the smallest radius. Such a profile leads to the injection of a huge amount of protons and a high $pp$ collision efficiency. Also, it seems that their 1\,GeV gamma-ray luminosity exceeds that of the kinetic luminosity of the wind at early stages. Our work takes into account the proton cooling due to inelastic collisions and adiabatic losses. Whereas the light curve in WLI decreases with time at early stages (see Fig.~2 in WLI), we expect that it is rather flat when the system is calorimetric in high-density regions and the injection of cosmic rays is supposed to be constant (see the previous section and discussions in \citealt{Lamastra17}). 
There is also a difference in the gamma-ray spectral shape between our result and that in WLI and WLII. In our calculation, the cutoff in the gamma-ray flux appears at a higher energy than that shown in WLI and WLII. The EBL cutoff should be around $10-20$~GeV, so the cutoff shown in WLI and WLII should not be caused by the EBL absorption. We do not show the neutrino flux since their the neutrino flux shown in WLII largely deviates from the theoretical expectation for the relationship between neutrino and gamma-ray fluxes. The gamma-ray and neutrino fluxes generated in $pp$ collisions should be roughly comparable at $E_\gamma\approx 2E_\nu$. The gamma-ray flux at 1\,GeV and all-flavor neutrino flux at 10\,TeV have a difference of $\approx(3/2)(20\,{\rm TeV}/1{\rm \,GeV})^{2-\Gamma}\simeq0.08$ for a SPL proton spectrum with $\Gamma_{\rm CR}=2.3$. This agrees with our result and in the result of \citet{Lamastra17}, while the result in WLII indicates that the neutrino flux at 10\,TeV is comparable to the gamma-ray flux at 1\,GeV.

The results of \citet{Lamastra17} are consistent with ours in terms of the spectral shapes of gamma-ray and neutrino emissions. However, the their fluxes themselves are about one order of magnitude larger than that of ours. Similarly to WLI and WLII, \citet{Lamastra17} also extrapolated an $R^{-2}$ profile for the gas density to very small radii, making the galactic disk a proton calorimeter. But since they considered the cooling of the accelerated protons due to $pp$ collisions, we do not see a large difference due to this extrapolation. On the other hand, all the accelerated protons are well confined and expand with the shock in our calculation, and the adiabatic cooling reduces the fluxes by a factor of $\sim 2$. In contrast, \citet{Lamastra17} assumed all the protons escape the shock and hence do not suffer from adiabatic losses. However, in reality, the escaping protons interact with the uncompressed gas with a smaller density, and they may also diffuse to a larger radius, where the gas density is very low. These effects are not considered in their calculation. If the diffusion coefficient is large, the $pp$ optical depth can be lowered by a factor of a few. On the other hand, if the diffusion coefficient is too small, the escaped protons could be caught up by the shock, implying that they cannot escape in the first place. 
Another important cause for the difference is that the shock expansion in the galaxy occurs for a short time. According to our calculation, the time that the forward shock experiences in the galactic disk is much shorter than the time in the halo. On the other hand, the total lifetime of AGNs (i.e., the Salpeter time) is about $\sim10^{7}\,$yr that is much shorter than the age of the galaxy. Thus, for the current observer, it is unlikely that the forward shocks in all the host galaxies throughout the universe are currently located in their galactic disks. To evaluate the contribution from all AGN-driven winds in the universe, it is more appropriate to firstly average the gamma-ray/neutrino luminosity over the entire evolution time and then sum up over redshifts, as done by WLI, WLII, and this work. Note that the gamma-ray and neutrino luminosities from the shock in the halo should be lower due to the much lower gas density, so this can further lower the diffusive flux by another factor of $3-5$. Lastly, \citet{Lamastra17} adopted a different luminosity function and redshift evolution of AGNs. 

{The employed luminosity function includes both radio-quiet AGNs and radio-loud AGNs. Radio-loud AGNs are accompanied by powerful jets, which may have contributions to the diffuse neutrino background \citep{Murase08nu, Murase13, BT14, Hooper16nu} by the interactions inside large-scale structures, or by interactions in the AGN core regions. Note that AGN winds and jets are not mutually exclusive. Our result does not exclude the possibility of these powerful jets as the sources of high-energy neutrinos.}

\section{Implications for the Diffuse Neutrino Background}
Obviously, the AGN-wind model has large model uncertainties, which are difficult to estimate. In principle, one could increase the neutrino flux to fit the IceCube's observation with extreme parameters, by, for example, assuming a larger ratio of the wind's kinetic luminosity to AGN's bolometric luminosity and a larger fraction of the thermal energy converted to non-thermal energy of the accelerated protons in the shock. However, the gamma-ray flux will then be increased by the same level, approximately matching the flux of EGB but overshooting the flux of IGRB \citep{Fermi15}. 
Indeed, a strong constraint on the sources of the cumulative neutrino background is unavoidable, as found by \citet{Murase13}. There are four facts: (1) the measured energy flux of neutrinos at $10-30\,$TeV is the comparable to that of IGRB at 100\,GeV, say, $10^{-7}\,\rm GeVcm^{-2}s^{-1}sr^{-1}$; (2) the gamma-ray flux and neutrino flux simultaneously produced via $pp$ collisions are comparable beyond 1\,GeV; (3) photons with energy above 100\,GeV will initiate electromagnetic cascades and form a diffuse gamma-ray background with an approximate $E^{-2}$ spectrum up to $\lesssim 100$\,GeV with an energy flux similar to the injected one, provided that they are injected at cosmological distances \citep{Strong73, Berezinsky75, Berezinsky11, Kachelriess12, Murase12}; (4) blazars, including the unresolved ones contribute $86_{-14}^{+16}\%$ of the EGB above 50\,GeV \citep{Fermi16} \citep[see also][]{Lisanti16} and $50_{-11}^{+12}\%$ below 10\,GeV \citep{Ajello15}, shrinking the room for the contribution of other sources to IGRB around 10\,GeV down to a level of at most 50\%. As a result, as shown by previous works, a strong tension against the IGRB data is unavoidable when such $pp$ scenarios explain the measured neutrino flux below $100\,$TeV \citep[e.g.,][]{Murase16,Chang16,Xiao16}, especially in the presence of cosmogenic gamma rays \citep{Murase16_PRD, Liu16, Berezinsky16}.
Note that the medium-energy component is unlikely to be Galactic. The latest shower and medium-energy starting event analyses suggest that the arrival distribution of neutrinos with these energies is consistent with an isotropic distribution \citep{IC17_ICRC}, and there is no evidence for a special source around the Galactic center, Fermi bubbles, and other structures such as Loop I. In addition, HAWC has already given a strong limit on the diffuse gamma-ray flux, $E_\gamma^2\Phi_\gamma\sim(0.3-1)\times{10}^{-7}~{\rm GeV}~{\rm cm}^{-2}~{\rm s}^{-1}~{\rm sr}^{-1}$, in the $30-100$ TeV range around the Galactic halo region \citep{HAWC17a, HAWC17b}. Such diffuse gamma-ray limits can constrain $pp$ scenarios, in which a significant fraction of IceCube neutrinos are explained by Galactic sources \citep{Ahlers14}. For example, if the Fermi bubbles~\citep{Fang17,Sherf17} or Loop I~\citep{Andersen17} dominantly contributes to the $10-100$ TeV neutrino flux, the diffuse gamma-ray flux from the sky region $\Delta\Omega$ is expected to be $E_\gamma^2\Phi_\gamma\sim4\times{10}^{-7}~{(3~{\rm sr}/\Delta\Omega)}~{\rm GeV}~{\rm cm}^{-2}~{\rm s}^{-1}~{\rm sr}^{-1}$, which is higher than the existing limits. Thus, the Galactic origin of these diffuse isotropic $10-100$ TeV neutrinos is unlikely.

Thus, we focus on the extragalactic interpretation of IceCube neutrinos. Taking the case of $\Gamma_{\rm CR}=2.3$ as an example, we perform the chi-square analysis and place a limit on AGN-driven wind models. For our model, we find that in order not to violate the 95\%~C.L. of the IGRB data at each energy bin\footnote{The one-sided 95\%~C.L. upper limit of the IGRB at the $i$th energy bin is calculated by $F_{\rm 95\%U.L.}(E_i)=F_{\rm IGRB}(E_i)+\sqrt{\chi_{90\%}^2}\sigma_{\rm IGRB}(E_i)$ where $F_{\rm IGRB}(E_i)$ is the measured IGRB flux at the $i$th energy bin and $\sigma_{\rm IGRB}(E_i)$ is the statistical error of the IGRB flux at the $i$th energy bin. $\chi_{90\%}^2=2.71$ is the chi-square value for 90\%C.L. for one degree of freedom.}, the gamma-ray flux can be shifted upwards by a factor of 7.3 at most {if the amplitude of the flux is taken to be a free parameter and the spectral shape is fixed}. The neutrino flux will be increased by the same factor resulting in a flux of $4.2\times 10^{-8}\,\rm GeVcm^{-2}s^{-1}sr^{-1}$ at 10\,TeV which is still about 5 times smaller than the best-fit flux of the IceCube neutrinos at 10\,TeV. This upper limit for our model template is shown with the blue curves with downward arrows in Fig.~\ref{fig:flux_compare}. {Considering the uncertainty in the measurement of IGRB due to the Galactic gamma-ray foreground, IGRB flux can only be increased by a factor of $\sim 1.5$ which only slightly changes the result here.} 
The spectral shape of diffuse gamma-ray flux obtained by \citet{Lamastra17} is very similar to ours, so the most constraining energy bin of the IGRB (the one in $4.5-6.4\,$GeV) should be the same with the one in our model. Therefore, the obtained upper limit for our model should also apply to the model of \citet{Lamastra17}. 
This is consistent with the multi-messenger constraints obtained by \citet{Murase13} for $pp$ scenarios, which showed that the spectral index cannot be softer than $\Gamma\sim2.1-2.2$ in order not to overshoot the IGRB data, when the neutrino flux is normalized to 100\,TeV. Note that the constraint becomes stricter if one normalizes the neutrino flux to the observation at 10\,TeV, ruling out even a harder slope. Indeed, for $\Gamma_{\rm CR}=2.1$, we find that the gamma-ray flux can be increased only by a factor of 2 at most in order not to violate the IGRB data, and the neutrino flux in this limit is still only $\sim30-40$\% of the best-fit flux at 10\,TeV. 
Although harder spectra would help to alleviate the tension, the observed neutrino spectrum in the $10-100$\,TeV range is too soft to explain with such hard spectra. This tension becomes severer if we recall that a large fraction of the IGRB is attributed to blazars. Thus, the AGN-driven winds can not be the dominant sources for $\gtrsim 10\,$TeV neutrinos, unless the sources are opaque to $10-100$\,GeV photons (which implies hidden cosmic-ray accelerators) \citep{Murase16}. 
However, since most gamma rays are emitted at $\lesssim 10$\,kpc according to Fig.~\ref{fig:lc}, it is unlikely that an intense soft X-ray photon field appear at this radius and effectively absorbs $10-100\,$GeV gamma-ray photons inside the host galaxy. 

{In addition, we note that only gamma rays from $\pi^0$ decay is considered as in the analysis of \citet{Murase13}. Given a hard proton spectrum $\Gamma_{\rm CR}\lesssim 2.1$, we can expect the GeV gamma-ray flux cascaded from higher energies to be comparable or even more important than that from $\pi^0$ decay at GeV energies if the internal absorption of gamma-ray is not much important (which is valid in the AGN-wind case). This can be seen from the $\Gamma_{\rm CR}=2.1$ case (blue curves) in Fig.~\ref{fig:flux_result}. Thus, even if only neutrinos above 100\,TeV are ascribed to certain species of extragalactic sources with a hard injection proton spectrum, the accompanying GeV gamma-ray flux would reach a level of $10^{-7}\,\rm GeVcm^{-2}s^{-1}sr^{-1}$ if there is little internal absorption of these gamma rays. The current analysis of the blazar contribution to the IGRB still has some uncertainties. If future measurements can lower the non-blazar component of the IGRB to half of the current level or even below, some of our model assumptions, including various other models for $\gtrsim 100\,$TeV neutrinos, can be tested critically.}

\section{Implications for Point-source Detection}
\begin{figure}[htbp]
\centering
\includegraphics[width=0.9\columnwidth]{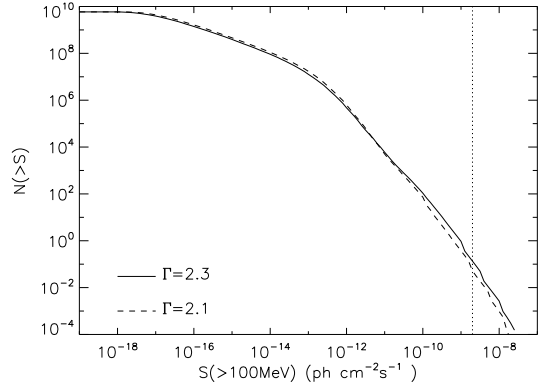}
\caption{Cumulative source count distribution of the gamma-ray flux above 100\,MeV from AGN-driven winds. The vertical dotted line is the 5$\sigma$ integrated sensitivity for a point source with a photon index of $\Gamma=2.0$.}\label{fig:CSD}
\end{figure}

AGN-driven winds are predicted to be persistent gamma-ray emitters, and current gamma-ray detectors such as \emph{Fermi}-LAT should be able to detect them with a  long-term exposure. The $5\sigma$ integrated sensitivity for 108 months of LAT observations on a high latitude point source above 100\,MeV with a photon index of -2.0 is\footnote{we obtain this sensitivity by multiplying a factor of $\sqrt{2/9}$ by the 2-yr sensitivity, which is given in {https://fermi.gsfc.nasa.gov/ssc/data/analysis/documentation/\\Cicerone/Cicerone\_LAT\_IRFs/
LAT\_sensitivity.html}.} $S_{\rm lim}(>100\,{\rm MeV})\approx10^{-9}\,\rm ph~cm^{-2}s^{-1}$. 
We calculate the cumulative source count distribution $N(>S)$ based on the cases of $\Gamma_{\rm CR}=2.1$ and $\Gamma_{\rm CR}=2.3$ shown in Fig.~\ref{fig:flux_result}, with $S$ here being the photon flux in unit of $\rm ph\,cm^{-2}s^{-1}$ above 100\,MeV from a certain source. The result is shown in Fig.~\ref{fig:CSD}. As can be seen, the number of detectable sources is about unity. A longer-time (e.g., 10\,yr) monitoring would help to discover point sources. 
If the AGN-driven winds give the dominant contribution to the IGRB, a few sources can be detected by \emph{Fermi}-LAT, which is consistent with the previous constraint on the effective source number density \citep{Murase16_PRD}. Our result is also consistent with the fact that a starburst co-existing AGN, NGC 1068, was detected by {\it Fermi}-LAT. 

Detecting individual neutrino sources with IceCube seems difficult. However, IceCube-Gen2 will be able to detect high-energy neutrino signals from most of the known astrophysical sources, including galaxies with AGN-driven winds. \citet{Murase16_PRD} investigated detection prospects for nearby high-density galaxies and argued that NGC~1068 is one of the promising target sources for IceCube-Gen2. 

\section{Summary}
In this work, we studied the diffuse gamma-ray and neutrino fluxes produced by AGN-driven winds. The expansion of the AGN-driven winds into the ISM gas and halo gas in the AGN's host galaxy will form a wind bubble, which can accelerate protons at the forward shock. Gamma rays and neutrinos are produced through the decay of pions which can be generated via inelastic $pp$ collisions between the accelerated protons and the shocked gas. We solve the dynamic evolution of the wind from the disk to the halo of the host galaxy. Based on the evolution model, taking into account some details such as proton cooling processes, we calculated the proton cooling and resulting pion production processes. The diffuse gamma-ray and neutrino fluxes are obtained by summing up all possible contributions from AGNs in the universe. We found that the generated gamma-ray flux can accounts for $\lesssim 30\%$ of the EGB flux around 50\,GeV. For $\Gamma_{\rm CR}=2.3$, the resulting neutrino flux is several times lower than or even more than one order of magnitude lower than the observed IceCube flux at 10\,TeV. Given that model uncertainties are large, the neutrino flux may be increased several times with optimistic (but somewhat extreme) model parameters. However, independent of details of the models, the IGRB data already rules out the possibility that the dominant fraction of IceCube neutrinos is accounted for in this model, for soft indices of $\Gamma\gtrsim2.2$. This conclusion can be strengthened if the contribution of unresolved blazars is taken into account. However, with hard spectral indices of $\Gamma\lesssim2.2$, it is still possible to explain the higher-energy component of the diffuse neutrino flux above $\sim0.1$ PeV. { Another potential constraint on the contribution of neutrino from AGN winds is the synchrotron radiation of co-produced secondary electrons in the shocks, which are supposed to give rise to multi-wavelength radiation at a similar flux level. The predicted multi-wavelength fluxes can be compared to the observation, and relevant physical quantities may be constrained. In addition, primary electrons can also be accelerated in the shocks and their radiation may provide additional information about and the parameters \citep{JiangYF10, WL15}}.  

In the AGN-driven wind model we consider, one should keep in mind that the starburst activity and AGN activity are not mutually exclusive. In cosmic-ray reservoir models, multiple classes of cosmic-ray accelerators are generally allowed, and supernovae, hypernovae, AGN-driven outflows, and possible weak jets from AGNs (although jets may be relevant for powerful radio-loud quasars) may contribute as accelerators of cosmic-rays in the 1-100 PeV range \citep{Tamborra14}. The model also predicts that a few nearby sources can be detected by {\it Fermi}-LAT, which seems consistent with the detection of NGC 1068. 

\begin{acknowledgements}
We thank Dr. Xiawei Wang and Prof. Walter Winter for helpful discussion, and Prof. Jeremiah Ostriker for his insightful comment. This work is supported by the 973 program under grant 2014CB845800, and the NSFC under grants 11625312 and 11273016.
The work of K.M. is supported by Alfred P. Sloan Foundation and NSF grant No. PHY-1620777. R.-Y.L. thanks Prof. Huirong Yan for her friendly hospitality during his visit in DESY. 
\end{acknowledgements}

\appendix
\section{Dynamic Evolution of the AGN-wind Bubble}
\subsection{The Relation between AGN's Bolometric Luminosity and Dark Matter Halo Mass}
One can check the details in \citet{WL15} and the references therein, we just follow and summarize their treatment here for convenience. AGN's bolometric luminosity is assumed to be a fraction $f_{\rm AGN}=0.5$ of the Eddington luminosity of the SMBH $L_{\rm Edd}=1.38\times 10^{38}(M_{\rm BH}/M_\odot)\rm \,erg/s$. The mass of the SMBH is related to the bulge stellar mass by \citep{McConnell13}
\begin{equation}
{\rm log}(M_{\rm BH}/M_\odot)=8.46+1.05{\rm log}\left[\frac{M_{\rm bulge}}{10^{11}M_\odot} \right].
\end{equation}
On the other hand, they adopt the bulge-to-total stellar mass ratio $B/T= M_{\rm bulge}/M_\star$ to be 0.3. Finally, one can obtain the dark matter halo mass through the relation \citep{Moster10}
\begin{equation}
M_\star=M_{\star,0}\frac{(M_{\rm halo}/M_1)^{\gamma_1}}{\left[1+ (M_{\rm halo}/M_1)^{\beta}\right]^{(\gamma_1-\gamma_2)/\beta}}
\end{equation}
where ${\rm log}(M_{\star,0}/M_\odot)=10.864$, ${\rm log}(M_1/M_\odot)=10.456$, $\gamma_1=7.17$, $\gamma_2=0.201$ and $\beta=0.557$. Thus, given the AGN's bolometric luminosity, one can find the halo mass by the above equations. Note that an upper limit for $M_{\rm halo}=10^{13}M_\odot$ is set, following WLI.

\subsection{Two-temperature Plasma Cooling in the Shocked Wind Region}
Following \citet{FG12}, we also consider the two-temperature effect in the plasma. In this scenario shock heats protons and electrons to different temperatures (Eq.~(\ref{eq:rh_tem_p}) and Eq.~(\ref{eq:rh_tem_e})), with an initial difference of $m_p/m_e$. With a higher initial temperature, the Coulomb collision between protons and electrons will transfer thermal energy from protons to electrons until $T_p=T_e$. The timescale to reach such a equilibrium reads \citep{FG12}
\begin{equation}\label{eq:equi_tem}
t_{\rm pe}=\frac{3m_em_p}{8(2\pi)^{1/2}n_{p,\rm sw}e^4\ln \Lambda}\left(\frac{kT_e}{m_e}+\frac{kT_p}{m_p} \right)^{3/2},
\end{equation}
with
\begin{equation}
\ln \Lambda \simeq 39+ \ln \left(\frac{T_e}{10^{10}\, \rm K}\right)-\frac{1}{2}\ln \left(\frac{n_{e,\rm sw}}{1\,\rm cm^{-3}}\right)
\end{equation}
where $n_{p,\rm sw}=n_{e,\rm sw}=n_{\rm sw}$ is the proton number density in the shocked wind region given. In \citet{FG12}, the authors adopt an analytic approximate proton cooling timescale by assuming Coulomb collision is the only process that changes the temperature of protons. However, as we mentioned in Section.~2, the adiabatic expansion and the freshly injected protons from the reverse shock will also effect the proton temperature. Thus, instead of adopting the oversimplified approximation, we trace the temperature of protons and electrons separately during the evolution. Note that, Eqs.~(\ref{eq:rh_tem_p}) and (\ref{eq:rh_tem_e}) is only valid for the temperature immediately after the shock. The average temperature in the entire downstream region is more relevant to the evolution. Therefore, we only use Eqs.~(\ref{eq:rh_tem_p}) and (\ref{eq:rh_tem_e}) as an initial temperature while calculate the proton temperature and electron temperature during the evolution by $T_p=P_{p,\rm sw}/n_{\rm sw}k$ and $T_e=P_{e,\rm sw}/n_{\rm sw}k$ respectively. The evolution of $P_{p,\rm sw}$ and $P_{e,\rm sw}$ can be found by Eq.~(\ref{eq:PErelation})$-$(\ref{eq:th_rate_e}). In Eq.~(\ref{eq:th_rate_e}), the Compton cooling/heating term $L_c$ is given by
\begin{equation}
L_c=-\frac{dU_{\rm ph}}{dt}V_{\rm sw}
\end{equation}
where $V_{\rm sw}=4\pi(R_s^3-R_{\rm rs}^3)/3$ and the rate of energy transfer between the electrons and photons owing to Compton scattering is given by \citep{Sazonov01}
\begin{equation}
\begin{split}
\frac{dU_{\rm ph}}{dt}=&4U_{\rm AGN}\sigma_Tn_{e,\rm sw}c\left(\frac{kT_e}{m_ec^2}-\frac{kT_X}{m_ec^2}\right)\left(1+\frac{5}{2}\frac{kT_e}{m_ec^2}-2\pi^2\frac{kT_X}{m_ec^2}\right)\\
&+4U_{\rm CMB}\sigma_Tn_{e,\rm sw}c\left(\frac{kT_e}{m_ec^2}-\frac{kT_{\rm CMB}}{m_ec^2}\right)\left(1+\frac{5}{2}\frac{kT_e}{m_ec^2}-2\pi^2\frac{kT_{\rm CMB}}{m_ec^2}\right).
\end{split}
\end{equation}
This formula is valid up to mildly relativistic electrons ($kT_e\leq 0.1~m_ec^2$). $T_X$ is the Compton temperature of the radiation field of AGNs, which is found to be in a narrow range around $2\times 10^7\,$K \citep{Sazonov04} and $\sigma_T$ is the Thomson cross section. $U_{\rm ph}=U_{\rm AGN}+U_{\rm CMB}=2.8\times 10^{-12}(L_b/10^{45}{\rm \,erg/s})(R/10{\rm \,kpc})^{-2} + 4\times 10^{-13}(1+z)^4\rm\, erg/cm^3$. Besides the Compton scattering, electrons can also lose energy via free-free emission and synchrotron radiation as considered in Eq.~\ref{eq:th_rate_e}. The cooling time due to free-free emission for electrons is $t_{\rm ff}=\frac{3n_ekT_e}{\epsilon_{\rm ff}}=4.7\times 10^8 \left(\frac{T_{e, \rm sw}}{10^{10}\rm \,K}\right)^{1/2}n_{e,\rm sw}^{-1}{g}_B^{-1}\,yr$, where $\epsilon_{\rm ff}=1.4\times 10^{-22}(T_e/10^{10}{\rm \,K}) n_e^2{g}_B\,\rm erg/cm^3/s$ is the thermal free-free emissivity with ${g}_B\sim 1$ being the average Gaunt factor. Synchrotron cooling time scale is $t_{\rm syn}=1.6\times 10^{12}\left(\frac{B}{1\mu G}\right)^{-2}\left(\frac{T_e}{10^{10}\rm \,K}\right)^{-1}\,$yr. We neglect the self-Compton scattering off the photons from free-free emission and synchrotron radiation here. 

\begin{figure}[htbp]
\centering
\includegraphics[width=1.0\textwidth]{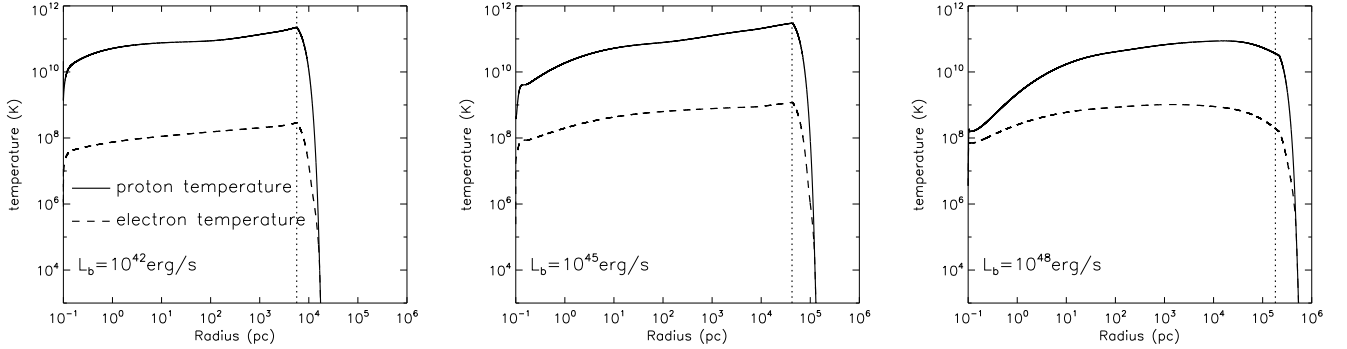}
\caption{Evolution of proton temperature (solid curve) and electron temperature (dashed curve) in the shocked wind region for $L_b=10^{42}\,$erg/s, $L_b=10^{45}\,$erg/s and $L_b=10^{48}\,$erg/s respectively. The vertical dotted lines show the Salpeter time after which the AGN shuts off.}\label{fig:tem_evol}
\end{figure}

In Fig.~\ref{fig:tem_evol}, we show the evolution of proton temperature and electron temperature in the case of $L_b=10^{42}\,$erg/s, $L_b=10^{45}\,$erg/s and $L_b=10^{48}\,$erg/s. Due to the continuous injection of freshly shocked wind, the temperature of proton and electron can not reach equilibrium before the AGN shuts off. The only exception is at very small radius when the wind has not been well decelerated as can be seen around $R=0.1\,$pc in the case of $L_b=10^{48}\,$erg/s (for $L_b=10^{42}\,$erg/s and $L_b=10^{45}\,$erg/s, this happens at a smaller radius than the initial radius considered in calculation). The reverse shock is very weak so that the temperature of shocked wind is low, resulting in a relaxation time between protons and electrons (i.e., $t_{pe}$) shorter than the dynamic timescale.

\subsection{Initial Condition}
\begin{figure}[htbp]
\centering
\includegraphics[width=0.7\columnwidth]{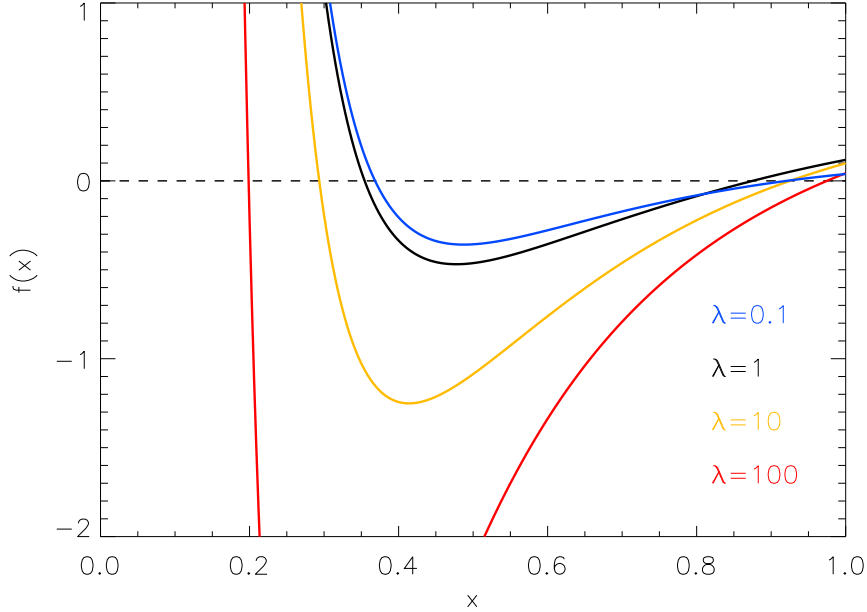}
\caption{Value of $f$ as a function of $x$. The solutions are the values of $x$ where $f(x)=0$.}\label{fig:solx}
\end{figure}

To solve Eqs.~(\ref{eq:mom})-(\ref{eq:th_rate_e}), we need to find the initial values for $v_s$, $P_{\rm sw}$($E_{\rm t,sw}$), $R_{\rm rs}$ and $v_{\rm rs}$, so that we can solve the dynamic evolution by Runge-Kutta method. At the very early stage, the gravity is negligible and the cooling of shocked gas is not strong to effect the dynamics. Thus, we can assume both forward shock and the reverse shock are adiabatic and the flow conserves both energy and momentum. These two conditions gives the equations
\begin{equation}\label{mom_con}
M_{\rm sw}v_w=M_{\rm sg}v_{s}+ \int_{R_{\rm rs}}^{R_s}4\pi n_{\rm sw}m_pv_{\rm sw}(R) R^2dR
\end{equation}
and
\begin{equation}\label{en_con}
\frac{1}{2}M_{\rm sw}v_{\rm w}^2=E_{k,\rm sw}+E_{t,\rm sw}+E_{k,\rm sg}+E_{t,\rm sg}=\frac{1}{2}\int_{R_{rs}}^{R_s} 4\pi n_{\rm sw}m_p R^2 v_{\rm sw}^2dR+\frac{3}{2}P_{\rm sw}V_{\rm sw}+\frac{1}{2}M_{\rm sg}v_s^2+\frac{1}{2}M_{\rm sg}v_s^2
\end{equation}
respectively. The integrals in Eq.~(\ref{mom_con}) calculates the total momentum of the shocked wind and the one in Eq.~(\ref{en_con}) calculates the total kinetic energy of the shocked wind. Substituting the expressions for $M_{\rm sw}$ and $n_{\rm sw}$ into above the equations, and defining $x=R_{rs}/R_s$, $y=v_s/v_w$, $\lambda=M_{\rm sg}/(\dot{M}R_s/v_w)$, we can reduce the above two equations to
\begin{equation}\label{mom_con_red}
\left(\frac{1}{y}-x\right)=\frac{4y}{x^2}(1-x)+\lambda y
\end{equation}
and 
\begin{equation}\label{en_con_red}
\frac{1}{2}\left(\frac{1}{y}-x\right)=\frac{2y^2}{x^2}\left(\frac{1}{x}-1\right)+\frac{1}{2}\left(\frac{1}{y}-x\right)\left(1-\frac{y}{x^2}\right)^2+\lambda y^2
\end{equation}
We can find the relation between $x$ and $y$ from Eq.~(\ref{mom_con_red})
\begin{equation}
y=\frac{\sqrt{x^2+4\left(4\frac{1-x}{x^2}+\lambda\right)}-x}{2\left(4\frac{1-x}{x^2}+\lambda\right)}
\end{equation}
We then can solve $x$ by substituting this relation into Eq.~(\ref{en_con_red}). Note that there is no analytic solution, but nevertheless we can solve the equation numerically by looking for the value of $x$ making the function $f\equiv \frac{2y^2}{x^2}\left(\frac{1}{x}-1\right)+\frac{1}{2}\left(\frac{1}{y}-x\right)\left(1-\frac{y}{x^2}\right)^2+\lambda y^2-\frac{1}{2}\left(\frac{1}{y}-x\right)=0$. Fig.~\ref{fig:solx} shows the value of $f$ as a function of $x$, under different values of $\lambda$. As is shown, $x$ has two solutions for each $\lambda$. The smaller one is an extraneous root of the equation as it results in $y>1$, so only the larger one is adopted in our calculation. We select a sufficiently small radius such as $R_{s,0}=0.1\,$pc as the initial point, and obtain the initial conditions for the dynamic evolution of the shocked ambient gas, say, $R_{rs,0}=xR_s$, $v_{s,0}=xv_w$, $v_{rs,0}=(4v_{s,0}/x^2-v_w)/3$, as well as $P_{sw,0}$ via Eq.~(\ref{eq:rh_P}).
  
\section{Cooling of the Accelerated Protons}
Actually, the exact solution for $N_p(E_p,r;R)$ can be obtained by solving the energy transport equation. But this would make the overall calculation extremely time-consuming. So we here make some approximation to take into account energy losses of cosmic-rays.

Firstly, let us only consider the influences of $pp$ collisions and neglect the adiabatic cooling for the moment. Assume that a constant number fraction $f$ of the initial protons undergoes the interaction in a given time step. Denoting the initial number of protons at energy $E_p$ by $N_0$, and we can obtain the number of protons that still have not interacted thus still at energy $E_p$ after $n$ time steps by $N_0(1-f)^{n}$. Bearing in mind the fact that the cross section of $pp$ collision depends weakly on proton energy beyond the threshold, and assuming that a proton loses half of its energy in one $pp$ collision given the inelasticity $\kappa=0.5$, we can know that protons with initial energy  $2^iE_p$ ($i\in \mathbb{N},i\leq n$) can cool to energy $E_p$ after $i$ times interactions. So the number of protons cooled to energy $E_p$ from higher energies can be given by $\sum\limits_{i=1}^{n}N_iC_n^i(1-f)^{n-i}f^i$ where $N_i$ is the initial number of protons at energy $2^iE_p$ and $C_n^i=n!/i!(n-i)!$ is the combinations of taking $i$ elements without repetition from a set of total $n$ elements. For a power-law spectrum with index $\Gamma_{\rm CR}$, we have $N_i\simeq 2^{i(1-\Gamma_{\rm CR})}N_0$. Provided a small enough time step, we can always have $(0<)f\ll 1$. The total number of protons of $E_p$ after $n$ time steps of injection can be given by
\begin{equation}
\begin{split}
N&=N_0(1-f)^n+\sum\limits_{i=1}^{n}2^{i(1-\Gamma_{\rm CR})}N_0C_n^i(1-f)^{n-i}f^i=N_0\sum\limits_{i=0}^{n}C_n^i(1-f)^{n-i}\left(2^{1-\Gamma_{\rm CR}}f\right)^i\\
&=N_0\left[1-(1-2^{1-\Gamma_{\rm CR}})f\right]^n.
\end{split}
\end{equation}
Defining $a\equiv\left[(1-2^{1-\Gamma_{\rm CR}})f\right]^{-1}$, we can write the above equation as
\begin{equation}
N=N_0\left[\left(1-\frac{1}{a}\right)^a\right]^{(1-2^{1-\Gamma_{\rm CR}})nf}
\end{equation}
Denoting the step size of time by $\delta t$, the fraction $f$ is the interaction rate of $pp$ collision in one step, i.e., $f=\sigma_{pp}n_{\rm sg}c\delta t$. To obtain a precise result, the time step should be as small as possible, i.e., $\delta t\to 0$, so we have $f\to 0$, and hence $a\to \infty$. Thus, we have 
\begin{equation}
N=N_0\lim\limits_{a\to \infty }\left[\left(1-\frac{1}{a}\right)^a\right]^{(1-2^{1-\Gamma_{\rm CR}}nf)}={\rm exp}\left[-(1-2^{1-\Gamma_{\rm CR}})nf\right]
\end{equation}
At a time $\Delta t$ after the initial injection, the total number of steps $n=\Delta t/\delta t$ and hence $\tau_{pp}\equiv nf=\sigma_{pp}n_{\rm sg}c\Delta t$ is the total number of collisions happening in this period. If the gas density is time dependent as is true in this work, we have $nf=\int \sigma_{pp}n_{\rm sg}cdt$. Now considering protons of energy $E_p$ are injected when the shock front is at $r$, the number of protons have energy $E_p$ when the shock propagates to $R$ can be given by
\begin{equation}
N(E_p,r,R)=N^{\rm inj}(E_p,r){\rm exp}\left[ -(1-2^{1-\Gamma_{\rm CR}})\sigma_{pp}(E_p)c \int_{t(r)}^{t(R)}n_{\rm sg}(r')dt\right]=N^{\rm inj}{\rm exp}\left[-(1-2^{1-\Gamma_{\rm CR}})\tau_{pp}(E_p,r,R)\right]
\end{equation}

Now we consider the effect of adiabatic cooling and neglect the cooling via $pp$ collision for the moment.
The spectrum evolution of protons injected at certain radius $r$ can be given by the energy transport equation,
\begin{equation}
 \frac{\partial N}{\partial t}+\frac{\partial}{\partial E_p}\left(\dot{\gamma}_{\rm ad}N\right)=0
\end{equation} 
$\dot{E}_{p,\rm ad}\simeq -v_s(r)E_p/r$ is the adiabatic cooling rate. Expanding the second term, we obtain 
\begin{equation}
\frac{\partial N}{\partial t}-\frac{v_s}{r}N-\frac{v_sE_p}{r}\frac{\partial N}{\partial E_p}=0
\end{equation}
We assume the proton spectrum is a power-law with index $\Gamma_{\rm CR}$, so we have $E_p(\partial N/\partial E_p)=-\Gamma_{\rm CR} N$. Then, the above equation can be written to
\begin{equation}
\frac{\partial N}{\partial t}=-(\Gamma_{\rm CR}-1)\frac{v_s}{r}N
\end{equation}
and we find
\begin{equation}
N(E_p,r,R)=N^{\rm inj}(E_p,r)\left(\frac{R}{r}\right)^{1-\Gamma_{\rm CR}}
\end{equation}
We define $\tau_{\rm ad}(r,R)\equiv \int_r^R vdt/r'=\int_r^R dr'/r'=\ln(R/r)$ so we can rewrite the above equation as $N(E_p,r,R)=N^{\rm inj}(E_p,r){\rm exp}\left[-(\Gamma_{\rm CR}-1)\tau_{\rm ad}(r,R)\right]$. 

Finally, we consider both the cooling process and have
\begin{equation}
N(E_p,r,R)=N^{\rm inj}(E_p,r){\rm exp}\left[-(1-2^{1-\Gamma_{\rm CR}})\tau_{pp}(E_p,r,R) - (\Gamma_{\rm CR}-1)\tau_{\rm ad}(r,R) \right].
\end{equation}
{Note that in some studies only protons injected within the cooling time are considered, and these protons are assumed to be uncooled completely, i.e., $N(E_p,r,R)=N^{\rm inj}(E_p,r)\theta (t_{\rm cool}-t(r,R))$ where $\theta$ is the Heaviside step function, $t_{\rm cool}$ is the cooling time of the proton of energy $E_p$ and $t(r,R)$ is the time in which the shock propagates from $r$ to $R$ \citep[e.g.][]{Torres04, Lacki13, Lamastra17}. Our method includes those cooled protons with $t_{\rm cool}<t(r,R)$ and also considers cooling of those protons that are recently injected with $t_{\rm cool}>t(r,R)$, although the difference is not significant.}

\bibliography{ms.bib}

\end{document}